%% file: main.tex
\documentclass[12pt]{article}
\usepackage{newtxtext,newtxmath}
\usepackage{graphicx}
\usepackage[letterpaper,margin=1in]{geometry}
\date{}

\makeatletter
\renewcommand{\fnum@figure}{\textbf{Figure \thefigure}}
\renewcommand{\fnum@table}{\textbf{Table \thetable}}
\makeatother
\usepackage{scicite}
\usepackage{url}
\usepackage{longtable}
\usepackage{framed}
\usepackage{CJKutf8}

\newcommand{\zh}[1]{\begin{CJK*}{UTF8}{gbsn}#1\end{CJK*}}

\def\scititle{
	The Growing Self-Reliance of Chinese Innovation
}
\title{\bfseries \boldmath \scititle}
\author{
	Ziyu~Chen$^{1,2}$,
	Christopher~Esposito$^{2,3,\ast}$\\
	\small$^{1}$University of Hong Kong\\
	\small$^{2}$University of California, Los Angeles\\
    \small$^{3}$Georgia Institute of Technology\\
	\small$^\ast$Corresponding author. Email: christopher.esposito@anderson.ucla.edu
}

\begin{document} 

\linespread{1.0}\selectfont
\maketitle
\thispagestyle{empty}

\vspace{2em}
\begin{center}
{\large\bfseries Abstract}
\end{center}
\vspace{0.5em}
\begin{quote}
\noindent U.S. policy increasingly seeks to slow China's technological rise by restricting its access to American science, on the assumption that Chinese innovation depends on U.S. science. Linking the full corpus of Chinese invention patents to the global scientific literature, we show that this dependence has fallen in recent years: the share of the China-produced science behind Chinese patents rose from 1\% in 2000 to 26\% in 2025, overtaking the U.S. share in 2021. As China's reliance on U.S.-produced science fades, policies restricting access fall out of alignment with the U.S.' actual strategic position.
\end{quote}

\vfill
\noindent\rule{0.35\linewidth}{0.4pt}\par
\vspace{0.4em}
{\footnotesize\noindent\textit{Acknowledgments.}~The authors thank Olav Sorenson for helpful discussions and comments on earlier versions of this manuscript and participants at the Atlanta Academy for Science and Innovation Policy for feedback.\par}

\linespread{1.5}\selectfont
\clearpage

\noindent

\subsection*{Introduction}

One of the most concerted---and potentially most consequential---objectives of contemporary U.S. science and innovation policy is to impede China's capacity to develop frontier technologies~\cite{chatterji_how_2025, kwon_dual_use_2026}. Export controls on scientific instruments and datasets~\cite{kwon_dual_use_2026}, restrictions on international collaboration~\cite{sira_act_2026}, and executive actions tightening academic exchange and researcher mobility~\cite{mervis_lawmakers_2025} have all been proposed or enacted in order to slow the diffusion of scientific knowledge from the U.S. to China. The effectiveness of these measures rests on a common assumption: that Chinese innovation is not self-reliant, but depends on science produced in the United States. However, that assumption has never been comprehensively tested.

Nor is that assumption necessarily correct. When Chinese inventors developed the copper-based cathode, for instance, they improved on the price and performance of batteries while mainly building on scientific knowledge that was already available in China: the concept of layered oxide structures and knowledge of copper's redox properties had already diffused to China decades earlier, and the critical new discovery---that copper could be used as a reversible redox center within an oxide structure---was made by scientists based in Beijing \cite{hu_layered_2017}. If this example is emblematic of a broader pattern, and Chinese inventions largely draw on Chinese science, then restrictions on China's access to U.S. science may not impede its innovation.

Yet attempts to systematically measure the dependence of Chinese innovation on U.S. science have been blocked by a lack of suitable data. The most comprehensive record of Chinese technological innovation is the corpus of patents filed at the China National Intellectual Property Administration (CNIPA). In principle, one could assess China's reliance on U.S. science by checking whether CNIPA patents cite U.S.-based research. However, CNIPA citations are institutionally biased toward Chinese-language papers and have limited coverage before 2010~\cite{wang_chinese_2025}. The common workaround is to restrict attention to the CNIPA patents also filed at the U.S. Patent and Trademark Office or European Patent Office, where citation practices are standardized~\cite{marx_reliance_2020}. Yet these patent families capture only 1\% of CNIPA patents. No existing approach, in short, yields an unbiased measure of China's dependence on U.S. science.

To address this limitation, we developed the first comprehensive measure of the dependence of China's innovation on international science. Leveraging Natural Language Processing and large-scale data on patents and scientific papers, we constructed a dataset linking all 6.7 million CNIPA invention patents produced in China to their likely worldwide scientific origins. This approach allowed us to quantify the extent to which Chinese innovation depends on scientific knowledge produced in the United States, in China, and in other countries and regions around the world, and to measure changes in those dependencies across time.

We began our analysis by translating the titles and abstracts of all CNIPA invention patents produced in China to English, and linked the resulting documents to the prior scientific literature based on the co-appearance of scientific phrases in the titles and abstracts of the patent and paper documents, using the method of \cite{arts_beyond_2025} and validated by \cite{arts_melluso_2026}. These steps yielded an edge list connecting 6.4 million CNIPA patents to 38 million OpenAlex scientific papers through 2.3 million shared scientific phrases (Figure~\ref{fig:main}A). By aggregating these links to the application years of patents and the countries-of-affiliation of papers, we measured the share of the scientific phrases in Chinese patents that are found on the papers produced by each country. These shares indicate the countries that supply the core scientific ideas underlying Chinese innovation (see Methods and Materials for details).

\subsection*{Results}

We found that Chinese technological innovation is increasingly self-reliant---meaning that the scientific phrases found on its patents increasingly appear in domestically-produced articles (Figure~\ref{fig:main}B). While in 2000, only 1\% of the patent-paper linked phrases appeared in domestically-produced articles, by 2025, that share had risen to 26\%. All the while, the share of such phases that appeared in U.S.-produced articles declined, dropping from 32\% in 2000 to 17\% in 2025. Moreover, since 2021, a greater share of the scientific phrases on Chinese patents have appeared in China-produced China-produced articles than in articles produced in the U.S.

These striking dynamics could be driven by a number of mechanisms. One that we are able to rule out is differences in the quality of U.S. and Chinese science. Looking at high-quality articles (those published in journals with Impact Factors exceeding 10), as of 2025, 27\% of the linked phrases in Chinese patents were found in domestically-produced articles, while only 22\% were found in articles produced in the U.S. (Fig. \ref{fig:supp-baseline-if}). Thus, for scientific ideas published in high-quality journals, China's innovation now appears to be more reliant on Chinese science than on science produced in the U.S.

We additionally found that China's rising self-reliance extends to detailed knowledge beyond what is captured by individual scientific phrases. We constructed combinations of the phrases found on Chinese patents (of length 2 up to length \textit{n}) and repeated our analysis for each such \textit{n-tuple}. For example, if a patent contained the phrases ``layered oxide structure'' and ``copper redox'', we created the 2-tuple ``layered oxide structure copper redox'', which describes the interdependence between the two concepts (Fig. \ref{fig:main}). We searched for such phrase combinations in the prior scientific literature, identifying highly-specific knowledge antecedents of Chinese innovations.

We found that Chinese self-reliance is rising rapidly even in the detailed areas of knowledge represented by higher-order phrase combinations (Fig. \ref{fig:combined_share_if1}). By 2025, 33\% of the 2-tuples on Chinese patents appeared in domestically-produced articles, while only 14\% appeared in articles produced in the U.S. The gap between the China-produced and U.S.-produced shares rises with combination length, with 41\% of combinations of length 4 or more appearing in China-produced articles and only 12\% in articles produced in the U.S. Thus, China's self-reliance is greatest in these highly-detailed areas of knowledge.

In addition to science of high quality and detail, we found that China's rising self-reliance extends to their higher-quality inventions. We collected a sample of CNIPA patents that are of above-average quality---those filed at both the CNIPA and USPTO---and measured the share of their phrases that appeared in prior Chinese and U.S.-produced papers (Fig. \ref{fig:supp-translation}A). Among those patents, the share of phrases appearing in China-produced articles rose from 1\% in 2000 to 23\% in 2025, a similar increase to that of the full sample of CNIPA patents. As of 2025, a greater share of China-produced CNIPA-USPTO patents contained phrases that appeared in Chinese publications than in publications produced in the U.S., again demonstrating China's rising self-reliance. 

The sample of cross-filed Chinese CNIPA-USPTO patents also enabled us to test whether idiomatic differences across countries could drive our main results. Because we used machine translation to convert CNIPA patents to English, idiomatic discrepencies could lead our method to overstate China's self-sufficiency. For example, Chinese scientists often refer to acetaminophen---the active ingredient in Tylenol---by its international name, paracetamol (\zh{扑热息痛}). Because the term paracetamol is less likely to appear in U.S.-produced articles than articles produced in China, it could bias our results. To address this possibility, we measured the share of scientific phrases in the CNIPA and USPTO versions of each patent family that appeared in prior China-based and U.S.-based scientific articles, while holding the underlying invention constant.

Our results indicate that idiomatic differences are unlikely to drive our main results. Of the CNIPA-USPTO patent families, a similar share of the phrases found on the CNIPA documents and the USPTO documents appeared in China-produced articles (23\% for the CNIPA documents, and 21\% for the USPTO documents; Fig. \ref{fig:supp-translation}). Thus, to the extent that they are resolved in USPTO patent documents, our method is robust to idiomatic differences across countries.

Finally, we discovered that the rising self-reliance of Chinese innovation is not confined to small or marginal Chinese organizations, but is also present in its leading technological firms and universities. We extracted the patents produced by 14 of China's leading organizations, ranging across telecommunications (Huawei), consumer electronics (Xiaomi), automobile design (BYD), utilities (State Grid), and technical universities (Tsinghua Univerity and Zhejiang University), and studied trends in their reliance on domestically-produced science. We found that, as of 2025, the phrases on the patents produced by each such organization were more likely to be found in China-produced articles than in articles produced in the U.S. (Fig.~\ref{fig:flagship-overview}).

\subsection*{Results Across Critical Technology Areas}

Our aggregate results may obscure important heterogeneity across fields, particularly in the technologies that governments treat as matters of national security. The United States' CHIPS and Science Act of 2022 directed the National Science Foundation to designate a set of key technology focus areas---among them artificial intelligence, semiconductors, quantum technology, and biotechnology---as priorities~\cite{chips_act_2022}. These technological areas area distinguished not only by their economic value, but also their strategic importance for national security~\cite{chatterji_how_2025,chips_act_2022}.

\begin{figure}
	\centering
	\includegraphics[width=\textwidth]{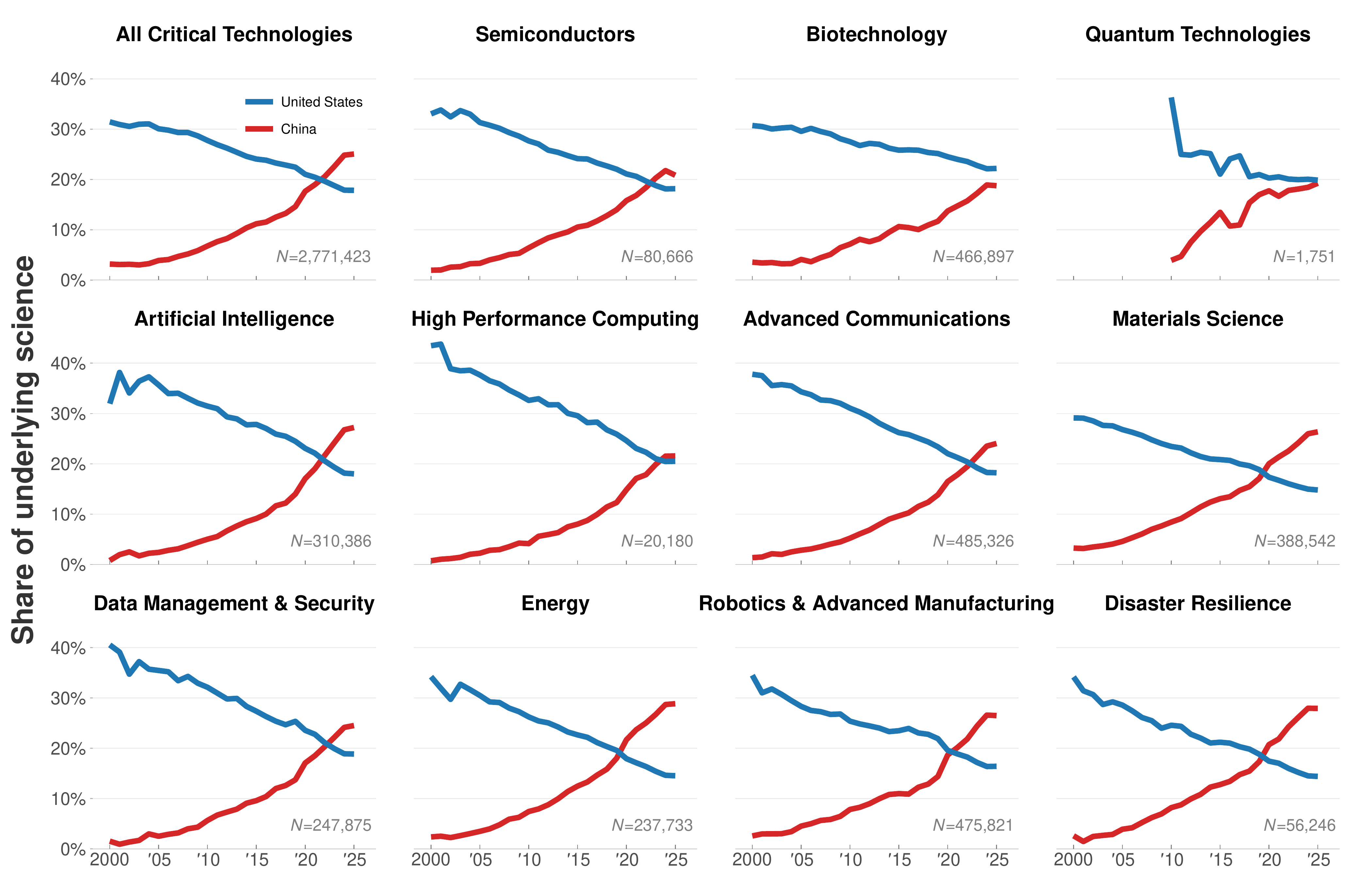}
	\caption{\textbf{Science underlying Chinese patents produced in the U.S. and China across Critical Technology Areas.} Each panel reports the share of linked scientific papers from the United States and China to CNIPA patents by a Critical Technology Area (CTA). Panel annotations report the number of CNIPA patents contributing to each area-specific estimate. See Methods and Materials for details on matching patents to CTAs.}
	\label{fig:cta-overview}
\end{figure}

We mapped CNIPA patents to Critical Technology Areas (CTAs; see Methods and Materials) and found that the self-reliance of Chinese innovation rose substantially across them (Figure~\ref{fig:cta-overview}). Across an aggregate grouping of CTAs, the share of phrases found on domestically-produced articles rose from 2\% in 2000 to 24\% in 2025, while the share found on U.S.-produced articles dropped from 31\% to 18\%. 

Across individual fields, a greater share of phrases on Chinese patents in Semiconductors, AI, Advanced Communications, Materials Science, Data Management \& Security, Energy, Robotics \& Advanced Manufacturing, and Disaster Resilience were more likely to be found on China-produced articles than than U.S.-produced ones as of 2025. Only in 3 fields (Biotechnology, Quantum Technology, and High Performance Computing) were such phrases more likely to be found on U.S.-produced articles, and even then these gaps were small and shrinking rapidly (Figure~\ref{fig:cta-overview}).

\subsection*{China's Origination of New Phrases}

The preceding results show that phrases on Chinese patents are now more likely to be found on articles produced in China than in the U.S. Those results, however, link patents only to where papers containing those phrases were produced, and not where the underlying phrases were originated. This distinction is important for policy, because if China primarily reproduces ideas that were originated in the United States, then restricting the flow of science could still impede Chinese innovation. On the other hand, if China originates these ideas itself, then no such leverage may exist. We therefore assigned each phrase to the first country in which it appeared on a publication and repeated our analysis.

A complication is that scientific knowledge is not all of the same vintage. Some phrases were introduced long ago and have since diffused widely, whereas a smaller, more recent set is at the knowledge frontier. These two sets of phrases carry different policy implications, so we examined them separately. For each phrase in each patent, we classified it as \textit{old} if it was introduced more than ten years before the patent's application, and as \textit{new} if it was introduced within ten years of the patent. 

\begin{figure}
	\centering
	\includegraphics[width=\textwidth]{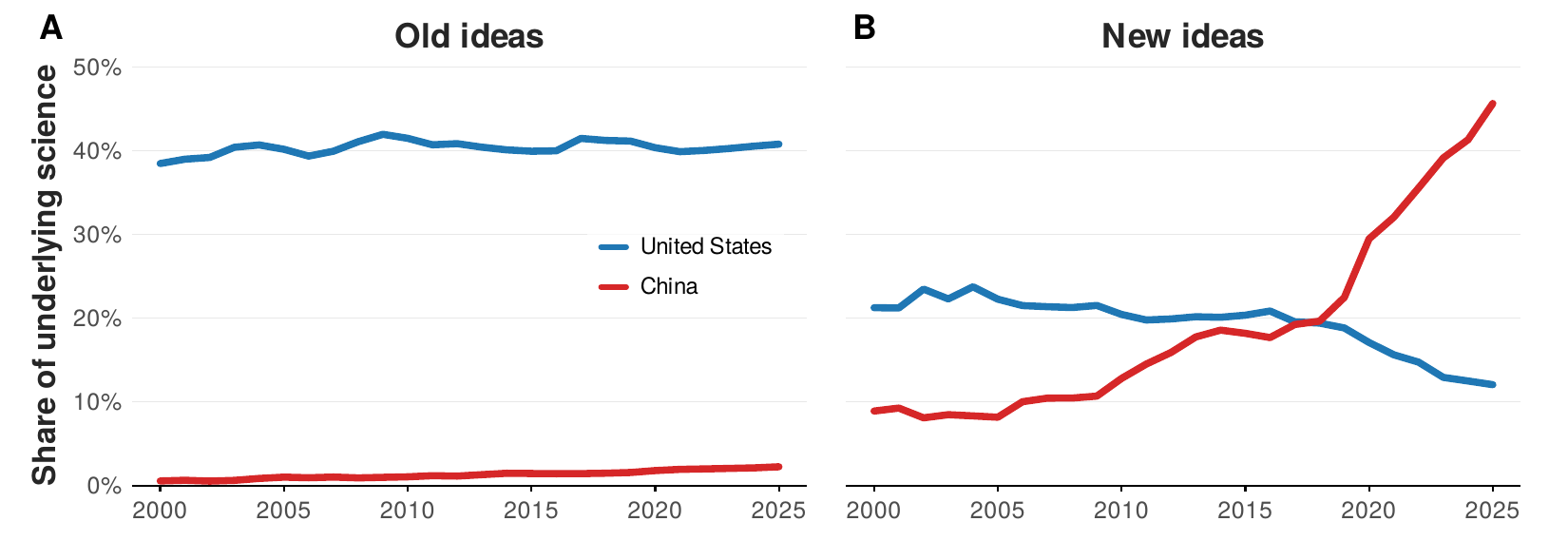}
	\caption{\textbf{Geographical origins of the scientific ideas underlying Chinese patents.} Each line shows the share of Chinese CNIPA patents originated by U.S.-based and China-based scientists. Phrases are split into two groups at the $Phrase*Patent$ level based on the age of the phrase (in calendar years) by the application year of a patent in which it appears. (\textbf{A}) Old ideas are phrases introduced $>$10 years the application year of a patent; (\textbf{B}) New ideas are phrases appearing on patents within 10 years of their introduction.}
	\label{fig:supp-rolling}
\end{figure}

We generated two key findings. First, China generally does not originate the old phrases used in its patents: as of 2025, 41\% of such phrases were originated in the U.S. and 3\% in China, and these shares have remained relatively stable over time (Fig. \ref{fig:supp-rolling}A). However, more importantly, the pattern reverses for newly-introduced phrases: the share of newly-introduced phrases that were originated in China rose from 7\% of the total to 46\% between 2000 and 2025, while the U.S.-originated share dropped from 21\% to 12\% (Fig. \ref{fig:supp-rolling}B).

Together these results identify two distinct channels through which China has expanded the self-reliance of its innovation. The first is the diffusion of older, foreign-originated ideas. Many of these ideas were introduced in the U.S. but have since diffused to China, and China now produces science using these ideas at scale. The second is the origination of new ideas within China. China is now the leading originator of the new ideas found in its patents, suggesting that has the capability to introduce the ideas its innovation will require in the future.

\subsection*{Discussion}

We have established, for the first time across the full corpus of Chinese invention patents, where the scientific ideas behind those inventions come from and how the geography of their sources have changed over time. Earlier efforts relied on patent citations, which are sparse, domestically-skewed, and non-representative. Linking patents to papers based on shared scientific phrases, though not without its own limitations, avoids these citation-related problems. 

We found that the self-reliance of China's innovation has risen sharply and broadly. The domestically-produced share of the science behind Chinese patents rose from 1\% in 2000 to 26\% in 2025, overtaking the U.S.-produced share in 2021. This is not an artifact of measurement---it holds when we limit the analysis to high-impact journals, when we analyze highly-precise idea combinations, when we examine China's largest firms and top universities, and when we infer a patent's scientific content from the English-language U.S. filing of the same invention. It is pronounced across nearly all of the technology areas that the U.S. government treats as strategically important, and it is driven by two core processes: China produces science built on older foreign-originated ideas at scale, and it originates a rapdily-growing share of the new phrases used in its own inventions.

These findings present challenges for U.S. policy. While China emerged as a scientific competitor to the U.S. largely by importing scientific knowledge and training its scientists at leading Western universities \cite{wu_chinas_2024}, such imports are losing their importance for sustaining China's continued advance. Policies aiming to slow China's innovation by curbing the export of scientific knowledge are thus premised on an outdated assessment of the U.S.' strategic position and are losing their effectiveness with time.

Isolationist policies also bear costs. By segmenting Chinese and U.S. science, knowledge export and collaboration restrictions slow scientific progress~\cite{kwon_dual_use_2026,alshebli_china_2023}, inhibit innovation \cite{esposito_global_2026}, and erode global norms of cooperation and reciprocity~\cite{chatterji_how_2025}. Most importantly, by fragmenting the co-development of critical technologies, such policies weaken opportunities for cross-monitoring and transparency, increasing the risk that technologies impacting global security could be developed and deployed without mutual oversight~\cite{wu_chinas_2024}. 

If leverage over China is sought, a strategy that the U.S. could pursue is to increase its own investments in science, with the intention of reinstating China's former dependence on the U.S. But while science investments would produce many benefits, it is unclear they would reinstate the former dependency: China now out-spends the U.S. in scientific research \cite{wagner_china_2026}, and its scientists are equally likely to lead China-U.S. cross-border scientific teams \cite{wu_chinas_2024}. Perhaps the U.S. could manage to pull ahead and reinstate a dependency, but sustaining that dependency would be another matter.

Eventually, U.S. policy will need to come to terms with the sophistication of China's internal innovative capabilities. Doing so may force the United States toward a China innovation policy that feels less than optimal. But a policy premised on China's actual domestic capabilities, even a suboptimal one, will serve the United States better than one premised on leverage it no longer holds.

\clearpage
\begin{figure}
	\centering
	\includegraphics[page=1,width=\textwidth]{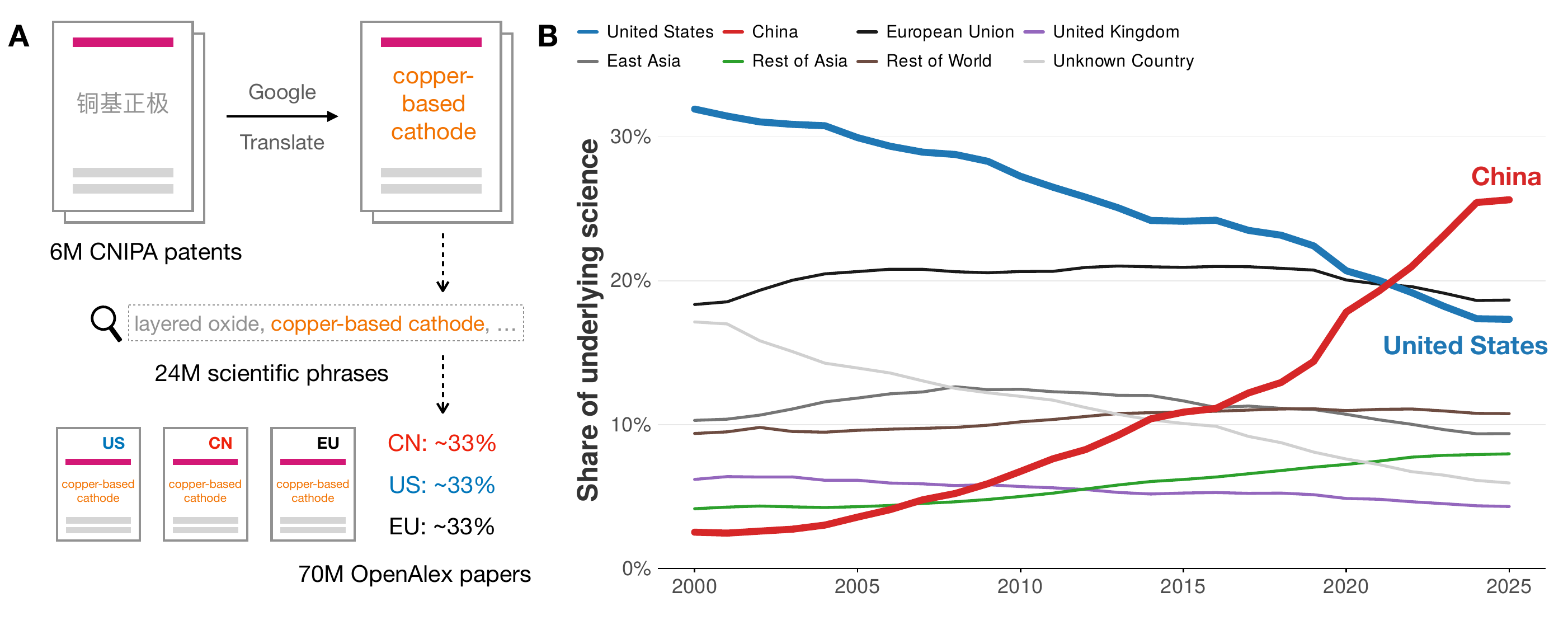}
	\caption{\textbf{Measuring the self-reliance of Chinese innovation.} (\textbf{A}) Schematic representation of the method used to link Chinese CNIPA patents to OpenAlex global scientific papers. CNIPA Patent abstracts and titles are translated into English using Google Translate. Their scientific phrases are extracted and matched to phrases appearing in the titles and abstracts of previously-published scientific papers. The resulting linkages are fractionally assigned the countries that produced the linked scientific articles to estimate each country's share of the scientific knowledge that is foundational to Chinese patents. (\textbf{B}) Changing producers of the scientific knowledge appearing in Chinese patents from 2000 to 2025, by patent application year. By 2025, Chinese CNIPA patents were more likely to connect to China-produced science than science from any other country.}
	\label{fig:main}
\end{figure}
\clearpage

\clearpage

\clearpage
\clearpage

\clearpage

\bibliography{bib_april_5_2026}
\bibliographystyle{sciencemag}

\newpage
\renewcommand{\thefigure}{S\arabic{figure}}
\renewcommand{\thetable}{S\arabic{table}}
\renewcommand{\theequation}{S\arabic{equation}}
\renewcommand{\thepage}{S\arabic{page}}
\setcounter{figure}{0}
\setcounter{table}{0}
\setcounter{equation}{0}
\setcounter{page}{1} 

\begin{center}
\section*{Supplementary Materials for\\ \scititle}
Ziyu~Chen,
Christopher~Esposito$^\ast$\\
\small$^\ast$Corresponding author. Email: christopher.esposito@anderson.ucla.edu\\
\end{center}
\subsubsection*{This PDF file includes:}
Materials and Methods\\
Supplementary Text\\
Figures S1 to S8\\
Tables S1 to S2
\newpage

\subsection*{Materials and Methods}

\subsubsection*{Data sources}

Our analysis draws on four data sources. First, we obtained the universe of CNIPA invention patents (including their titles, abstracts, application years, and International Patent Classification codes) from Google Patents, which supplies English titles and abstracts rendered by machine translation via Google Translate. Second, we obtained the scientific literature against which these patents are linked, namely paper titles, abstracts, publication years, and authorship affiliations, from the January 9, 2024 snapshot of OpenAlex. Importantly, this snapshot was taken before Elsevier abstracts were removed from OpenAlex, which happened in November of 2024. To obtain the country of origin for each publication, we extracted the affiliation country of the last author, as recorded in OpenAlex's authorship object. Third, the scientific phrase dictionary used to identify candidate phrases in both patents and papers is from~\cite{arts_beyond_2025}. Fourth, we obtained patent family information from PATSTAT, which we used to study CNIPA--USPTO patent families. Finally, we determined the country whether a CNIPA patent was China-produced based on whether it was applied for first at the CNIPA before other international patent offices.

The type of patents we study---invention patents---are one of the three types of patents issued by the CNIPA, which includes design patents and utility patents. At the CNIPA, utility patents protect the \textit{structure} of technologies as opposed to their function of claims. The patents we analyze---CNIPA invention patents---are most analogous to USPTO utility patents.

To ensure that we focused our analysis on reputable scientific articles, we restricted our sample of OpenAlex publications to those published in journals with 2021 Impact Factors of 1 or greater. In the supplement, we replicated our analyses after restricting the sample to papers in journals with Impact Factors greater than 5 and 10.

\begin{figure}
	\centering
	\includegraphics[width=0.82\textwidth]{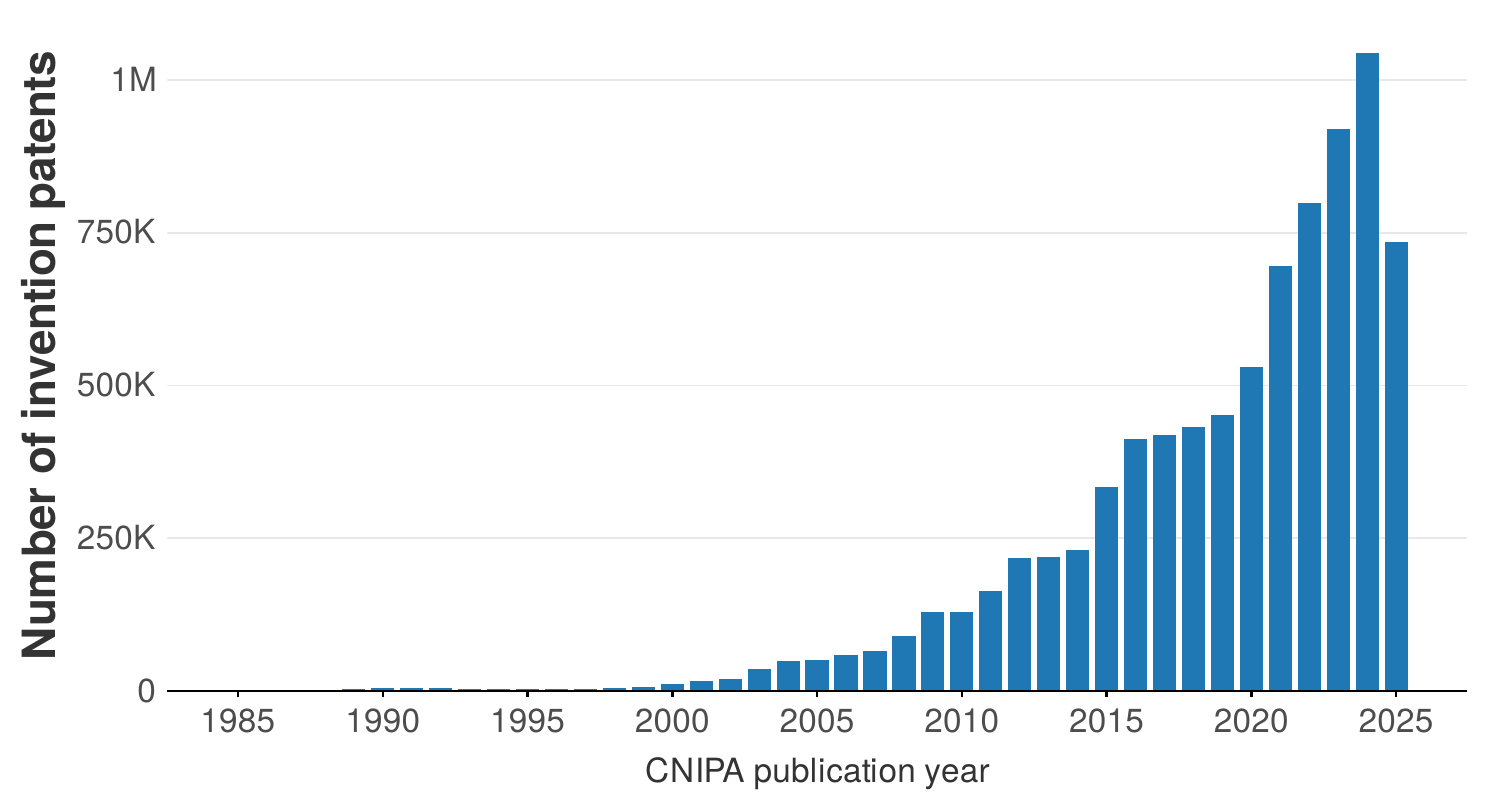}
	\caption{\textbf{CNIPA China-produced invention patents by publication year.} Counts of CNIPA China-produced invention patents used to construct the patent-paper linkage data, by publication year.}
	\label{fig:supp-patent-counts}
\end{figure}

Figure \ref{fig:supp-patent-counts} shows the number of China-produced CNIPA invention patents by application year that linked to at least one previously-published OpenAlex publication via a shared scientific phrase. The count of such patents is very low at the start of the data panel, but begins to climb in 2000, breaks 250,000 patents in 2015, exceeds one million patents in 2024, and drops in 2025 due to partial coverage for that year. 

\begin{figure}
	\centering
	\includegraphics[width=0.82\textwidth]{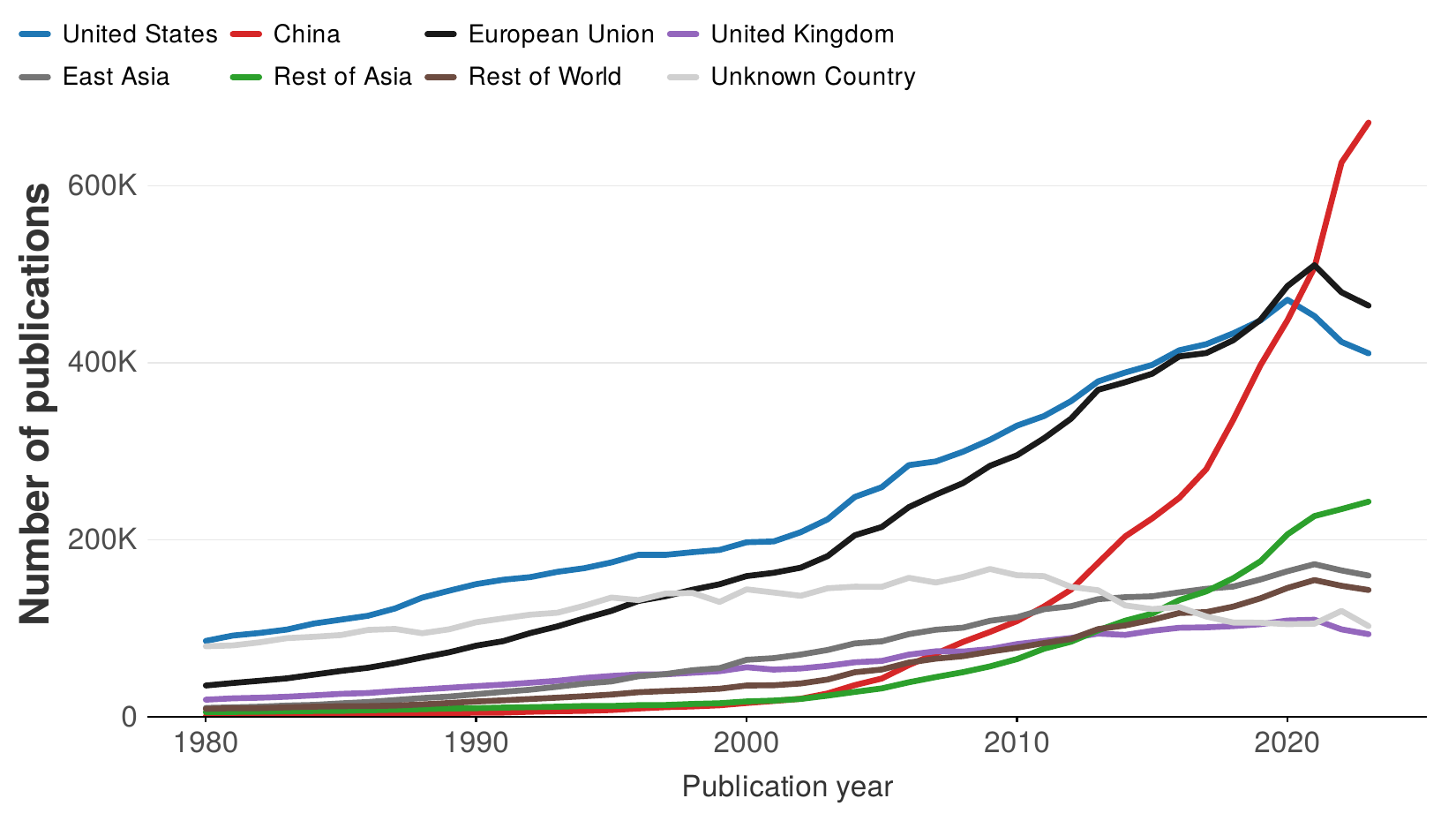}
	\caption{\textbf{Publications by geographical region and publication year.} Counts of scientific publications used to construct the patent-paper linkage data, by geographical region and publication year.}
	\label{fig:supp-publication-counts}
\end{figure}

Figure \ref{fig:supp-publication-counts} shows the number of OpenAlex papers produced by each country in journals with $Impact Factor \geq$ 1. This gives the number of papers that could be linked to patents via scientific phrases. The U.S. produced the largest count until 2018, after which it was eclipsed by the European Union and then China. China was the leading producer of scientific publications at the end of the data panel in 2025.

\subsubsection*{Linking patents and papers via shared scientific phrases}

We constructed links between patents and papers using three steps (Figure~\ref{fig:main}A).

First, we extracted the core scientific phrases from each patent our dataset using the dictionary developed by~\cite{arts_beyond_2025}. This dictionary contains over 27 million phrases drawn from 75 million scientific publications published between 1900 and 2023. The dictionary was developed by identifying the new terms appearing in scientific articles following a "burn-in" period of papers published before 1900. Therefore, phrases in the abstract and titles appearing for the first time after 1900 are considered to be scientific phrases and may appear in our dataset. 

Second, for each scientific phrase in each patent, we identified all OpenAlex papers containing the same phrase in their title or abstract and published at least one year before the patent's application year. To prevent ubiquitous, low-information phrases from dominating the linkage, we weighted each patent--paper edge by the inverse of the number of candidate papers containing the phrase (i.e., an edge weight of $1/c$, where $c$ is the count of candidate papers). The procedure yields an edge list linking 6.4 million patents to 38 million scientific publications through 2.3 million shared phrases.

Finally, we aggregated these weighted edges to the level of the patent application year and the country of the linked publications.

The validity of the method we use to link patents to papers has been studied extensively. Papers introducing such scientific phrases have been shown to predict external measures of scientific impact, such as the likelihood that a paper will be recognized with a Nobel Prize~\cite{arts_beyond_2025}, and phrase linkages between patents and papers were found to strongly correlate with survey responses from R\&D managers on their reliance on scientific ideas in their labs' technologies---even when the resulting patents made no citations to the scientific literature. Thus, phrase linkages between patents and papers appear to offer a meaningful pathway to identifying the scientific antecedents of technologies.

In addition to our main analysis involving independent scientific phrases, in the supplement we analyzed combinations of phrases. To do so, for each patent we formed n-tuples of its extracted phrases, enumerating all phrase combinations of length $k$, ranging from 1 up to length $n$. These n-tuples represent both the component-level and the combinatorial knowledge embodied in the invention ~\cite{fleming_technology_2001} (see Fig. \ref{fig:main} and the explanation in the main text).

\begin{figure}
	\centering
	\includegraphics[width=0.82\textwidth]{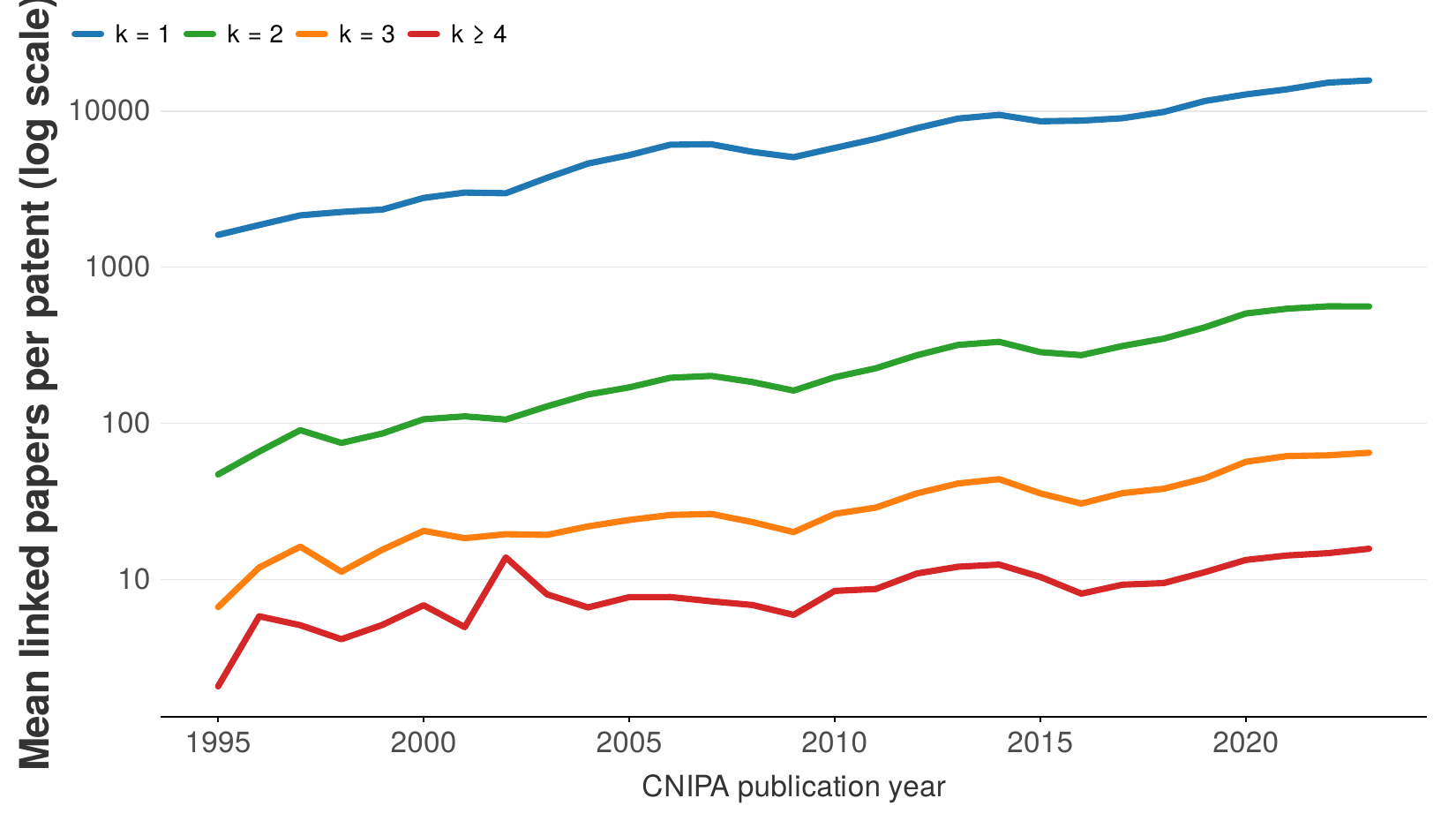}
	\\[1em]
	\caption{\textbf{Count of papers linked to patents over time.} Patents are linked to more papers using lower-order n-tuples (i.e. $k=1$ versus higher-order n-tuples (i.e. $k\geq4$), indicating that the use of higher-order n-tuples leads us to draw patent-paper links between more specialized knowledge units. 
    }
	\label{fig:supp-phrase-counts}
\end{figure}

Figure \ref{fig:supp-phrase-counts} shows the mean number of papers linked to each CNIPA patent, by patent grant year. The figure breaks out the data by the size of the n-tuple making the link, where $k=1$ designates the case where phrases are treated independently. A key finding produced by Figure \ref{fig:supp-phrase-counts} is that lower-order n-tuples (i.e. when $k=1$ link to more papers) link to more papers than do higher-order n-tuples (i.e. when $k>=4)$. For example, 1-tuples on patents granted in 2025 linked to, on average, over 10,000 prior papers, while the average tuple of length 4 or more ($k>=4$) linked on average to 11 prior papers. This finding supports our interpretation of low-order tuples as indicators of general knowledge, and high-order tuples as indicators of specialized knowledge.

\begin{figure}
	\centering
	\includegraphics[width=\textwidth]{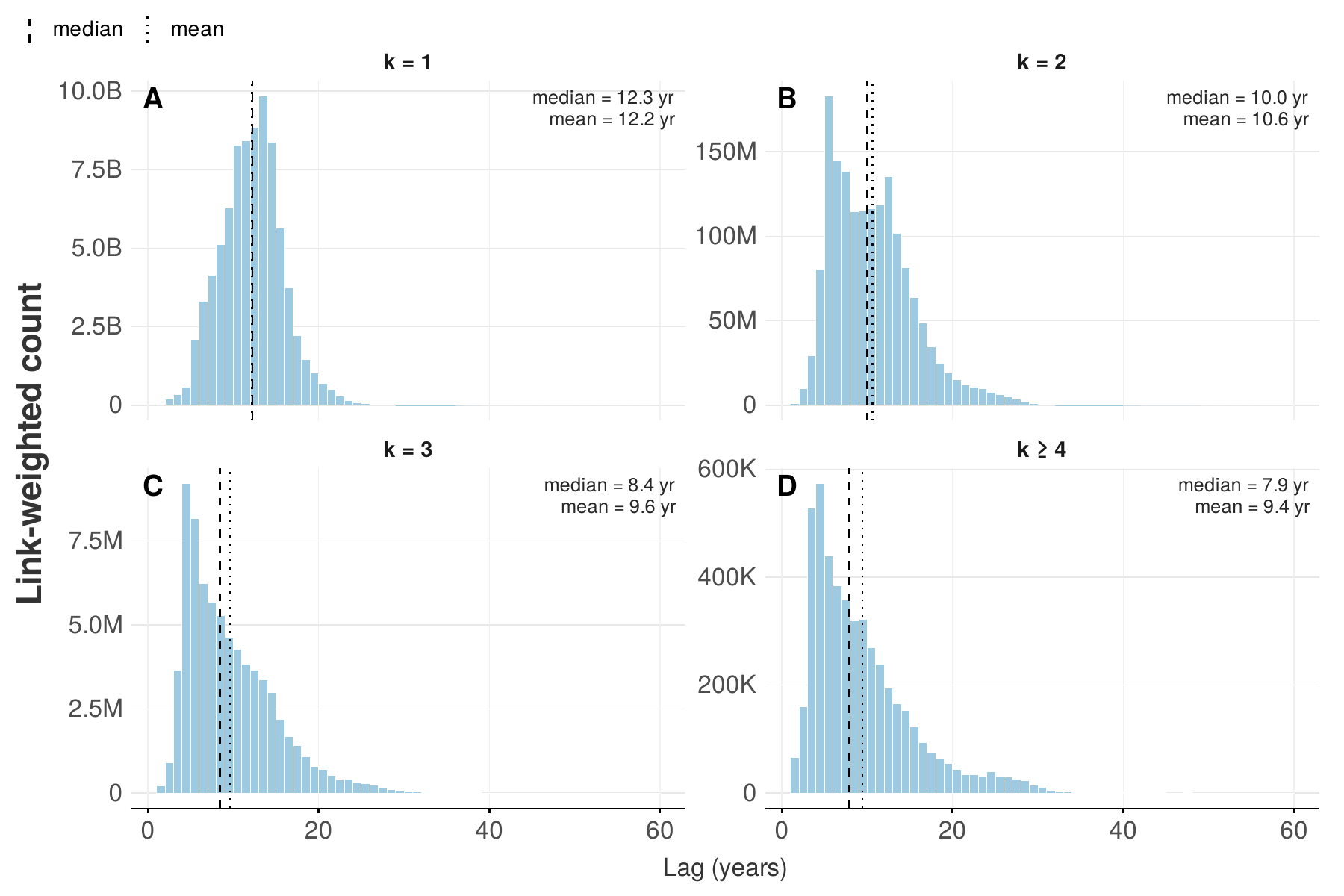}
	\caption{\textbf{Time lag of knowledge dependencies (T, weighted).} Full distributions of patent grant year minus publication year for the weighted T specification. (\textbf{A}) 1-tuples, (\textbf{B}) 2-tuples, (\textbf{C}) 3-tuples, (\textbf{D}) tuples of size 4 or larger.}
	\label{fig:supp-time-lag-t-weighted}
\end{figure}

While lower-order tuples tend to connect to more papers than higher-order tuples, they also tend to link to older papers. Figure \ref{fig:supp-time-lag-t-weighted} shows the distribution of time lags between the publication year of papers and the grant year of their linked patents, by tuple size. The time lag of links drawn using 1-tuples is roughly normally-distributed with a median length of 12.3 years (Fig. \ref{fig:supp-time-lag-t-weighted}A). As tuple size increases, the time lag distribution shifts inward and forms a long right-tail; for example, tuples of size 4 or greater have a median time lag between paper and patent of 7.9 years, but their long right-tail indicates that some such links have time lags of 20 years or longer (Fig. \ref{fig:supp-time-lag-t-weighted}D). These findings reinforce our interpretation of lower-order tuples as indicators of general knowledge and higher-order order tuples as indicators of specialized knowledge: while general scientific knowledge diffuses broadly across across technology over long periods of time, specialized knowledge diffuses relatively fast but with a few slow-to-diffuse applications in niche technological areas.

\subsubsection*{Linking CNIPA patents to critical technology areas}

The CHIPS and Science Act defined 10 technological for focused federal attention due to their importance for international security, future innovative growth, and relevance to national priorities. \cite{chips_act_2022}. We obtained from the NSF a list of keywords assigned to 11 critical technology areas---the additional area was added by the NSF splitting "Semiconductors" apart from its parent field of "High Performance Computing".

To measure scientific dependence within each of the 11 NSF-supplied Critical Technology Area (CTA), we mapped CNIPA patents to CTAs through their International Patent Classification (IPC) codes, the classification system applied to all Chinese patents. For each of the 7,666 IPC main groups, we used a large language model to determine whether the group's technical scope corresponded to one of the eleven NSF CTAs, supplying the model with the keyword lists that NSF developed (see Figure~\ref{fig:cta-prompt} for the classification prompt). The LLM returned a best-matching area together with a confidence score. This procedure classified 1,469 IPC groups into a critical technology area and judged the remainder non-applicable; Table~\ref{tab:cta-ipc-crosswalk} lists the full crosswalk of IPC main groups to critical technology areas. We then assigned each patent to an area on the basis of its primary IPC code, yielding area-specific patent populations ranging from 2,236 patents in Quantum Technologies to 648,100 in Advanced Communications. We also produced an aggregate group, labeled "All Critical Technologies", which includes all patents assigned to one of the CTAs. 

\begin{figure}
	\centering
	\includegraphics[width=\textwidth]{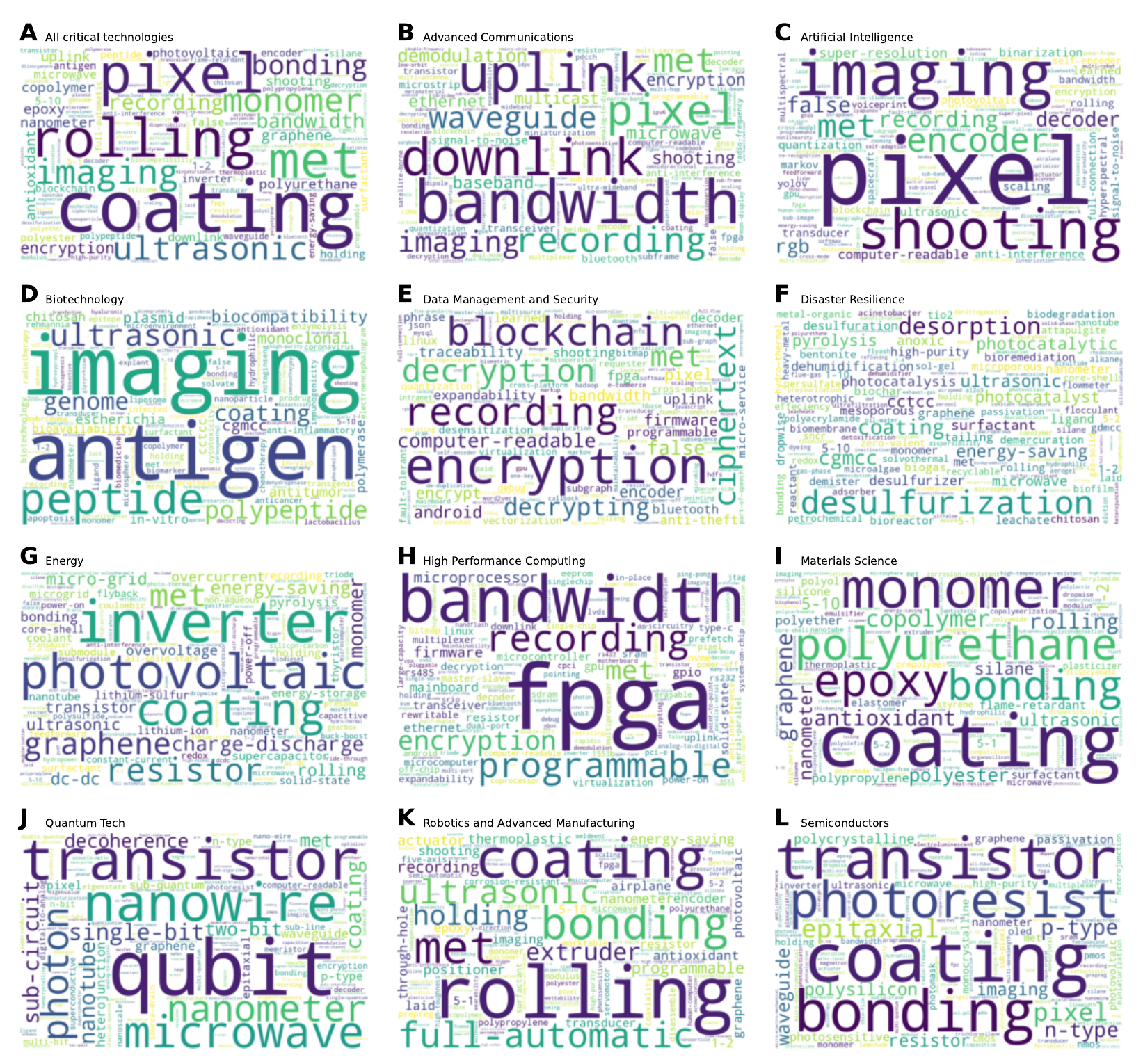}
	\caption{\textbf{Word clouds for linked one-token phrases in critical technology areas.} (\textbf{A}) All critical technologies. (\textbf{B}) Advanced communications. (\textbf{C}) Artificial intelligence. (\textbf{D}) Biotechnology. (\textbf{E}) Data management and security. (\textbf{F}) Disaster resilience. (\textbf{G}) Energy. (\textbf{H}) High performance computing. (\textbf{I}) Materials science. (\textbf{J}) Quantum technologies. (\textbf{K}) Robotics and advanced manufacturing. (\textbf{L}) Semiconductors.}
	\label{fig:supp-wordclouds}
\end{figure}

Figure \ref{fig:supp-wordclouds} shows word clouds of the scientific phrases on linked patents and papers, broken out by CTA. The major terms appearing in the word clouds for each CTA generally correspond to our expectations; for example, in Advanced Communications, the major terms are ``bandwidth'' and ``uplink'', in Data Management and Security, the major terms are ``blockchain'' and ``encryption'', and in Quantum Technology, the major terms are ``transistory'', ``nanowire'', and ``qubit''.

\subsubsection*{Linking CNIPA patents to Chinese organizations}

To study how self-reliance varies across individual Chinese organizations, we identified the CNIPA patents filed by fourteen prominent Chinese companies and universities. We took the assignee (applicant) names directly from Google Patents in their original Chinese-language form. We relied on the original-language names rather than Google's romanized, harmonized assignee field because the harmonized field is populated for fewer than one in five CNIPA patents and is essentially absent after 2015, whereas Chinese-language assignee names are available for virtually every CNIPA patent.

We selected the fourteen organizations as a purposive sample stratified across four types of Chinese innovators: private information-and-communications-technology firms (Huawei, Tencent, and ZTE); consumer-hardware and advanced-manufacturing firms (BOE, Xiaomi, OPPO, vivo, BYD, CATL, Gree, and Midea); a state-owned utility (State Grid); and elite public universities (Tsinghua University and Zhejiang University). Our selection of organizations is intended to be demonstrative rather than exhaustive or representative.

We then attributed patents to organizations by searching the assignee field for a set of curated character strings representing the organizations and their subsidiaries, listed in full in Table~\ref{tab:cn-org-matching}. Because Chinese organization names are compositional, a single distinctive stem usually captures a parent organization together with all of its subsidiaries and regional units at once: the stem \zh{华为} (``Huawei''), for example, matches the headquarters \zh{华为技术有限公司} as well as entities such as \zh{华为终端有限公司} and \zh{华为云计算技术有限公司}, without our having to enumerate them. We modified this simple rule in three ways. First, a few economically important subsidiaries share no characters with their parent and were added explicitly; the most consequential is Huawei's semiconductor subsidiary HiSilicon (\zh{海思半导体}, \zh{海思光电}). Second, because the search operates on substrings, some stems collide with unrelated organizations that happen to contain the same characters, so we specified exclusion strings to remove them: for example, the stem \zh{京东方} (``BOE'') is a substring of unrelated Beijing and Nanjing firms of the form \zh{北京东方}\ldots\ (such as the waterproofing manufacturer \zh{北京东方雨虹}), and the State Grid stem \zh{国网} is a substring of \zh{中国网}\ldots\ (such as the carrier China Netcom, \zh{中国网络通信}); both were excluded. Third, for organizations recorded partly in Latin script---most importantly OPPO, which appears as ``Oppo''---we additionally matched the romanized brand token with a case-insensitive search. A patent was attributed to an organization if any of its (possibly several) co-assignees matched that organization's strings.

Applying these patterns identified 508,293 CNIPA invention patents across the fourteen organizations, ranging from about 5,000 for CATL, to about 81,000 at Huawei, to more than 100,000 for State Grid (Table~\ref{tab:cn-org-matching}). Manual inspection of the matched names confirmed that residual false positives---small, unrelated firms whose names coincidentally contain a brand stem that is also a common word (for example \zh{小米}, ``millet'')---account for well under 0.1\% of the patents attributed to any organization.

\subsubsection*{Identification and analysis of first instances of scientific phrases}

To analyze the countries that originate the phrases underlying Chinese patents, we identified the first paper to use each such phrase in either its title or abstract. When a phrase first appeared in two or more papers in the same year, we selected one of those papers at random.

We further distinguished between old and new phrases, based on the number of years elapsed between the publication year of the paper containing the first instance of a phrase and the application year of a coincident patent. We defined "old" phrases as those introduced 10 or more years before the coincident patent, and "new" phrases as those appearing on patents within 10 years of their introduction. 

Importantly, the same phrase can (and often does) appear in multiple patents. In this case, the phrase may appear in patents at different ages, and possibly will appear as an ``old'' phrase in some patents and as a ``new'' phrase in others. 

Figure \ref{fig:supp-first-instances-lag} shows the age distribution of the time gap between the first instance of new phrases in publication records and their appearance in patent documents, broken out by n-tuple size. Phrases take several decades from the year they first appear in scientific publications until they diffuse broadly across patenting, with an median time lag between first introduction and the maximal number of patents published containing the phrase in a year of 42.7 for individual phrases (1-tuples; Fig. \ref{fig:supp-first-instances-lag}A). The corresponding lag is shorter for higher-order tuples.

\subsection*{Supplementary Text}

\subsubsection*{Robustness across thresholds for science quality}

Millions of scientific papers are indexed by OpenAlex---the complete OpenAlex repository contains over 240 million ``works'', which includes documents ranging from peer-reviewed publications in international journals, to government reports, to research documents of dubious quality. Not all of these works contain the sort of rigorous scientific knowledge that could be used to power innovation, so it is important to evaluate the quality of the underlying scientific research when attributing ideas embedded in patent documents to papers.

The method we use to link patents to papers naturally focuses on papers on higher quality science. The primary reason for this is that we draw links between patents and papers if they share common scientific phrases. Because a phrase needs to be relevant to both the patent and scientific literature in order to be used to make such a connection, our method generally focuses on the scientific articles that are relevant to technology.

In addition to this natural floor limit on paper quality, we also assessed the robustness of our results when filtering out papers in low-impact journals. In our main analyses, we restricted our sample to papers in journals with 2021 Impact Factors greater than or equal to 1. This restricts our main analysis to the top 21,658 journals and generally limits us to papers in journals with international interest.

\begin{figure}
	\centering
	\includegraphics[width=\textwidth]{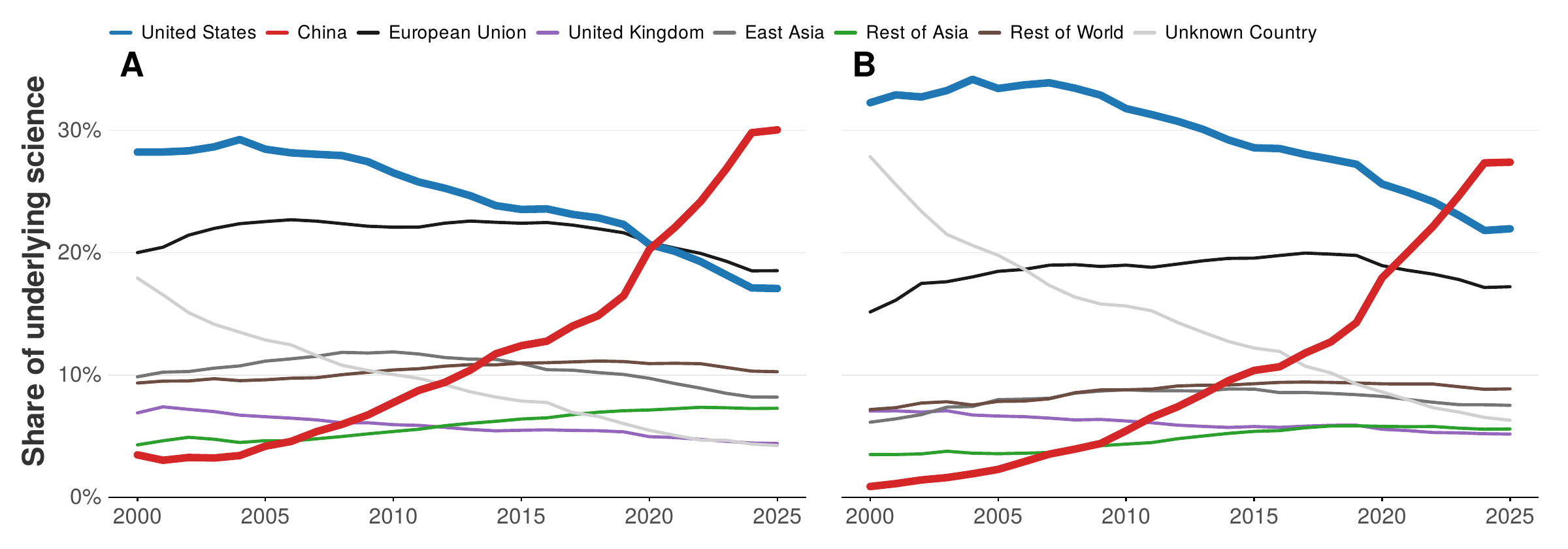}
	\caption{\textbf{Country shares of science underlying Chinese patents by Impact Factor thresholds.} (\textbf{A}) Papers in journals with 2021 Impact Factors $\geq$ 5. (\textbf{B}) Papers in journals with 2021 Impact Factors $\geq$ 10.}
	\label{fig:supp-baseline-if}
\end{figure}

\begin{figure}
	\centering
	\includegraphics[width=\textwidth,height=0.85\textheight,keepaspectratio]{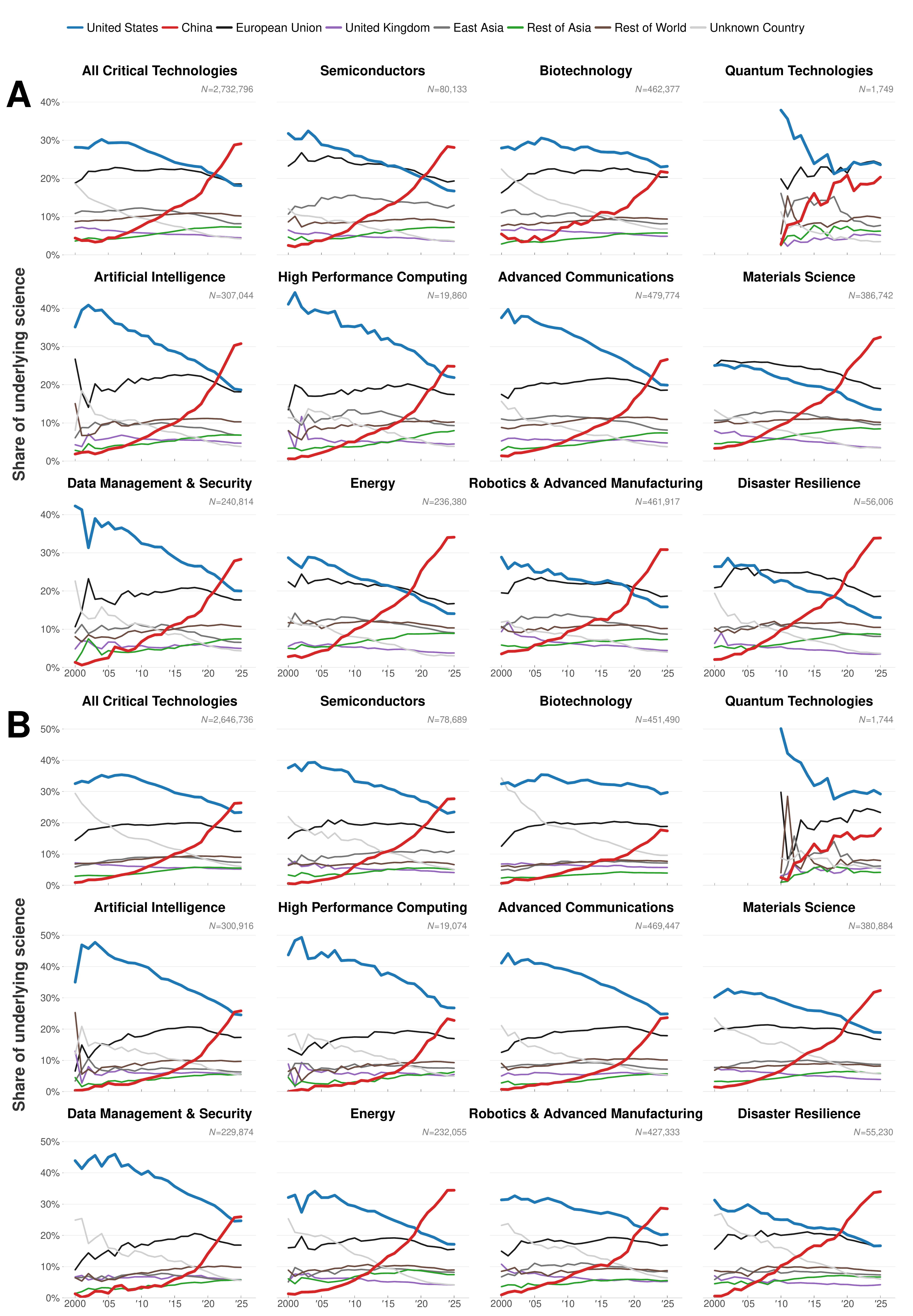}
	\caption{\textbf{Country shares of science underlying Chinese patents by Critical Technology Areas, broken out by journal 2021 Impact Factor.} (\textbf{A}) Papers in journals with 2021 Impact Factors $\geq$ 5. (\textbf{B}) Papers in journals with 2021 Impact Factors $\geq$ 10.}
	\label{fig:supp-cta-overview-if}
\end{figure}

In addition to our main results, we also tested the robustness of our findings using more stringent Impact Factor thresholds. We found similar results across these robustness checks: Figure \ref{fig:supp-baseline-if} shows that the China share of underlying science exceeded that of the U.S. share in 2025 under both the Impact Factor $\geq$ 5 and Impact Factor $\geq$ 10 thresholds. Meanwhile, for Critical Technology Areas, Figure \ref{fig:supp-cta-overview-if} shows that in 2025 the China share exceeded the U.S. share of underlying science in the aggregate ``All Critical Technologies'' grouping under both such thresholds. In addition, the 2025 China share lagged behind the U.S. share in a similar set of CTAs under these higher Impact Factor thresholds as it did in our main analysis, such as Biotechnology---confirming that the field-specific patterns of China's self-reliance is robust to higher Impact Factor thresholds. Finally, Figure \ref{fig:supp-old-new-if} shows that the rising propensity for China to originate the new knowledge underlying its patents---and the lack of growth in its propensity to originate the underlying old knowledge---extends across the higher Impact Factor thresholds, with similar results obtained using the Impact Factor $\geq$ 5 and Impact Factor $\geq$ 10 thresholds as in the main analysis.

\subsubsection*{Extended discussion on the origination of phrases in China and the rest of the world}

\begin{figure}
	\centering
	\includegraphics[width=\textwidth]{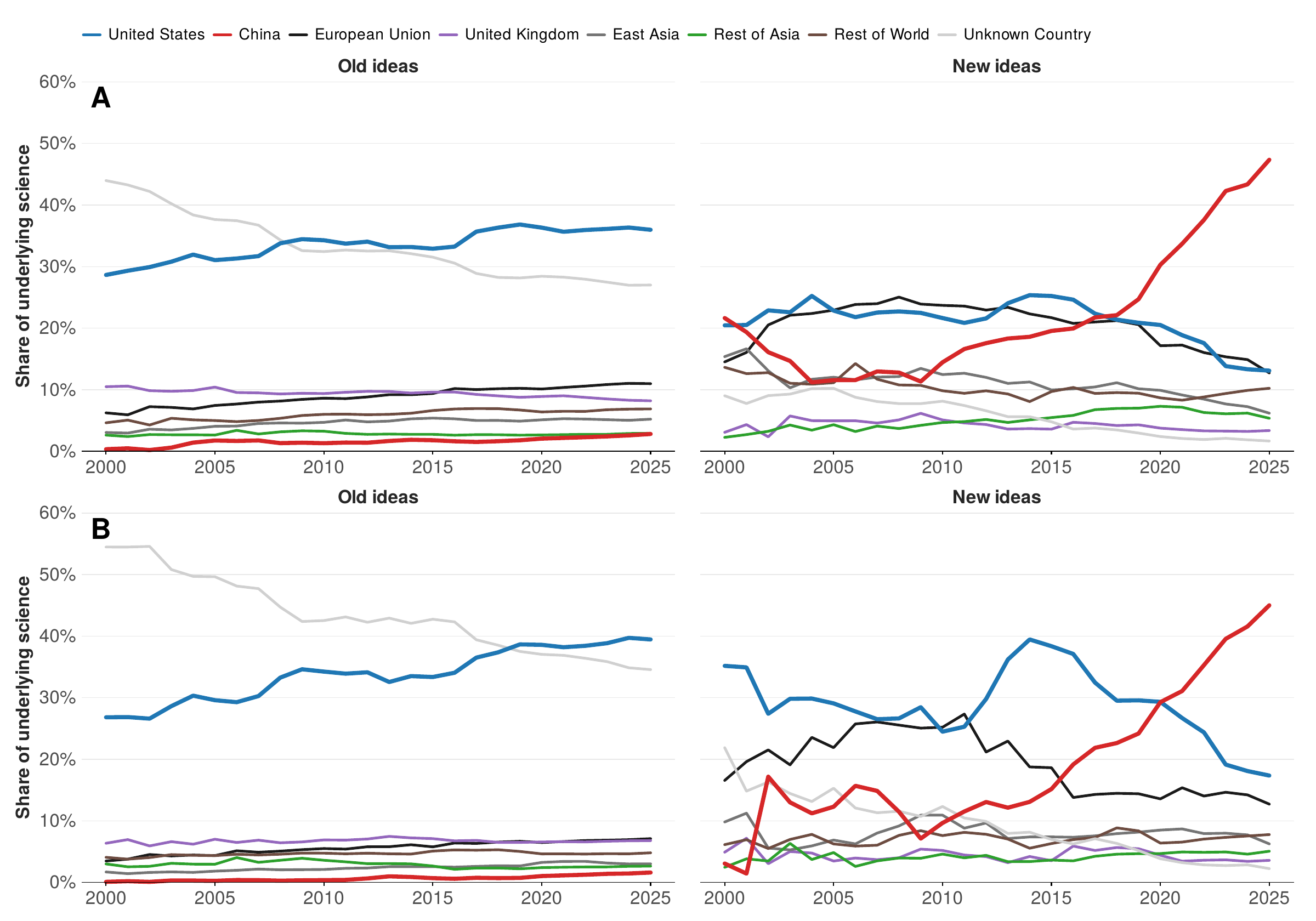}
	\caption{\textbf{Country shares of underlying science origination by country, for old and new, broken out by journal 2021 Impact Factor.} (\textbf{A}) Papers in journals with 2021 Impact Factors $\geq$ 5. (\textbf{B}) Papers in journals with 2021 Impact Factors $\geq$ 10.}
	\label{fig:supp-old-new-if}
\end{figure}

We use the first appearance of each scientific phrase to determine where it originated. The U.S. remained the leading originator of the \textit{old} phrases used in Chinese patents throughout the study period, while China overtook the U.S. as the leading originator of \textit{new} phrases in 2017 (Fig.~\ref{fig:supp-rolling}); these dynamics are robust to restricting attention to high-impact journals (Fig.~\ref{fig:supp-old-new-if}).

\begin{figure}
	\centering
	\includegraphics[width=\textwidth]{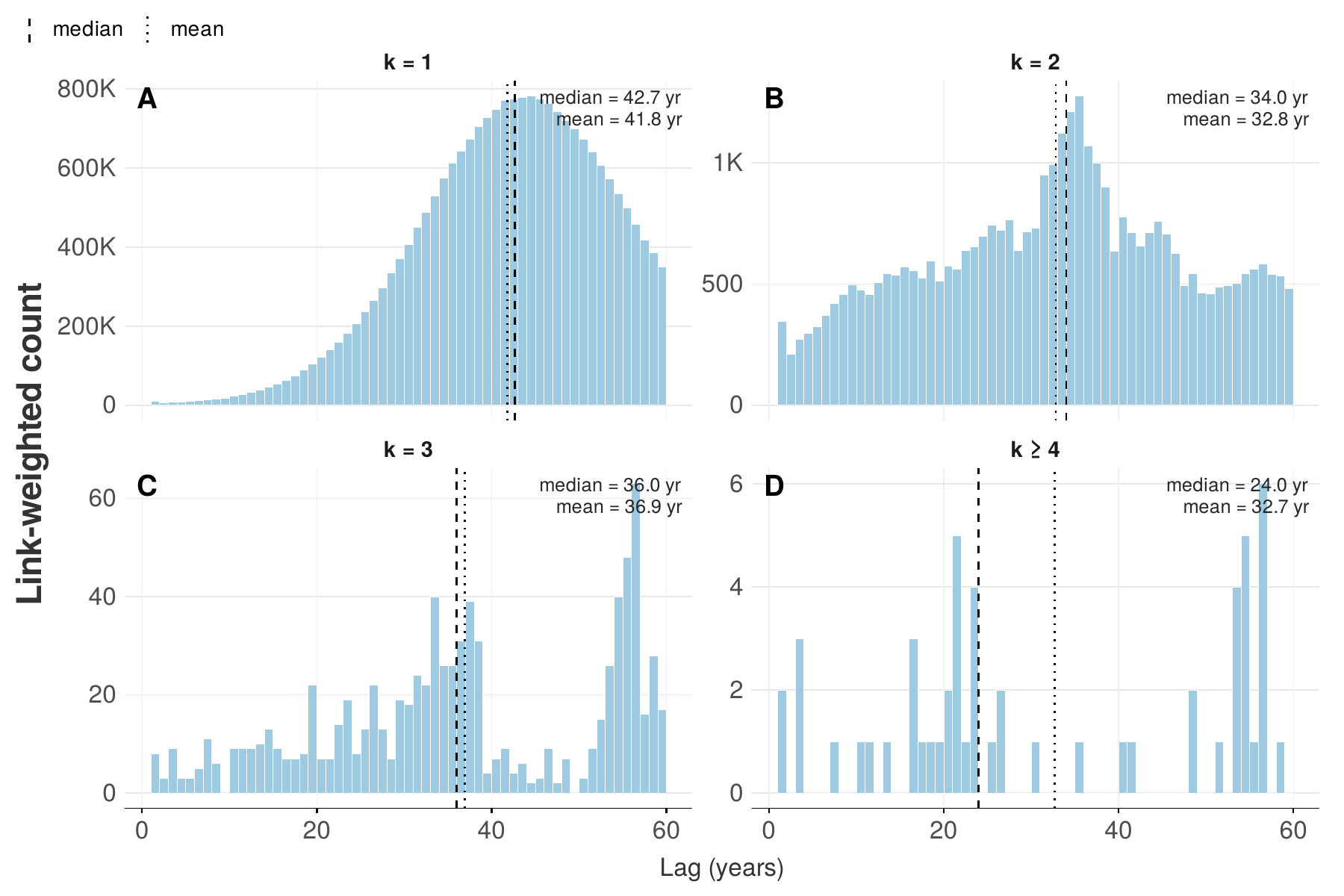}
	\caption{\textbf{Time lag between phrase introduction and appearance in patent documents by n-tuple size.} (\textbf{A}) 1-tuples, (\textbf{B}) 2-tuples, (\textbf{C}) 3-tuples, (\textbf{D}) tuples of size 4 or larger.}
	\label{fig:supp-first-instances-lag}
\end{figure}

Interpreting these results requires one further observation: the vast majority of phrases in Chinese patents are old, with a median lag of 42.7 years between a phrase's origination and its appearance in a patent (Fig.~\ref{fig:supp-first-instances-lag}). Since most phrases are old, and the U.S. originated far more of these old phrases than China did even through 2025 (Fig.~\ref{fig:supp-rolling}), China's rising overall self-reliance can be reconciled only one way: China now produces science containing these old, foreign-originated phrases at scale---in larger quantities than the U.S. itself. This is the principal mechanism behind China's growing reliance on domestically produced knowledge (Fig.~\ref{fig:main}).

The new ideas underlying Chinese innovation follow the opposite pattern. They are now far more likely to be introduced by China-based than U.S.-based scientists (Fig.~\ref{fig:supp-rolling}). Though they constitute a much smaller share of the phrases embedded in Chinese patents (Fig.~\ref{fig:supp-first-instances-lag}), they are the ones China would not otherwise possess: were they originating abroad, export restrictions could plausibly curb Chinese innovation. However, because new phrases now originate within China, knowledge export restrictions cannot affect them. Thus, for old and new ideas alike, Chinese innovation is approaching self-reliance.

That China is becoming self-reliant while still building heavily on U.S.-originated phrases (Fig.~\ref{fig:supp-rolling}) reveals the lasting influence of U.S. science on the Chinese innovation ecosystem. China may now produce its new ideas domestically and hold ready access to older ones---but those older ideas were nonetheless developed abroad. Self-reliance is not the same as independence from this history: while Chinese innovation is growing more self-reliant, the true \textit{indigeneity} of Chinese innovation appears much further off.

\newpage



\clearpage

\begin{figure}
	\centering
	\includegraphics[width=\textwidth]{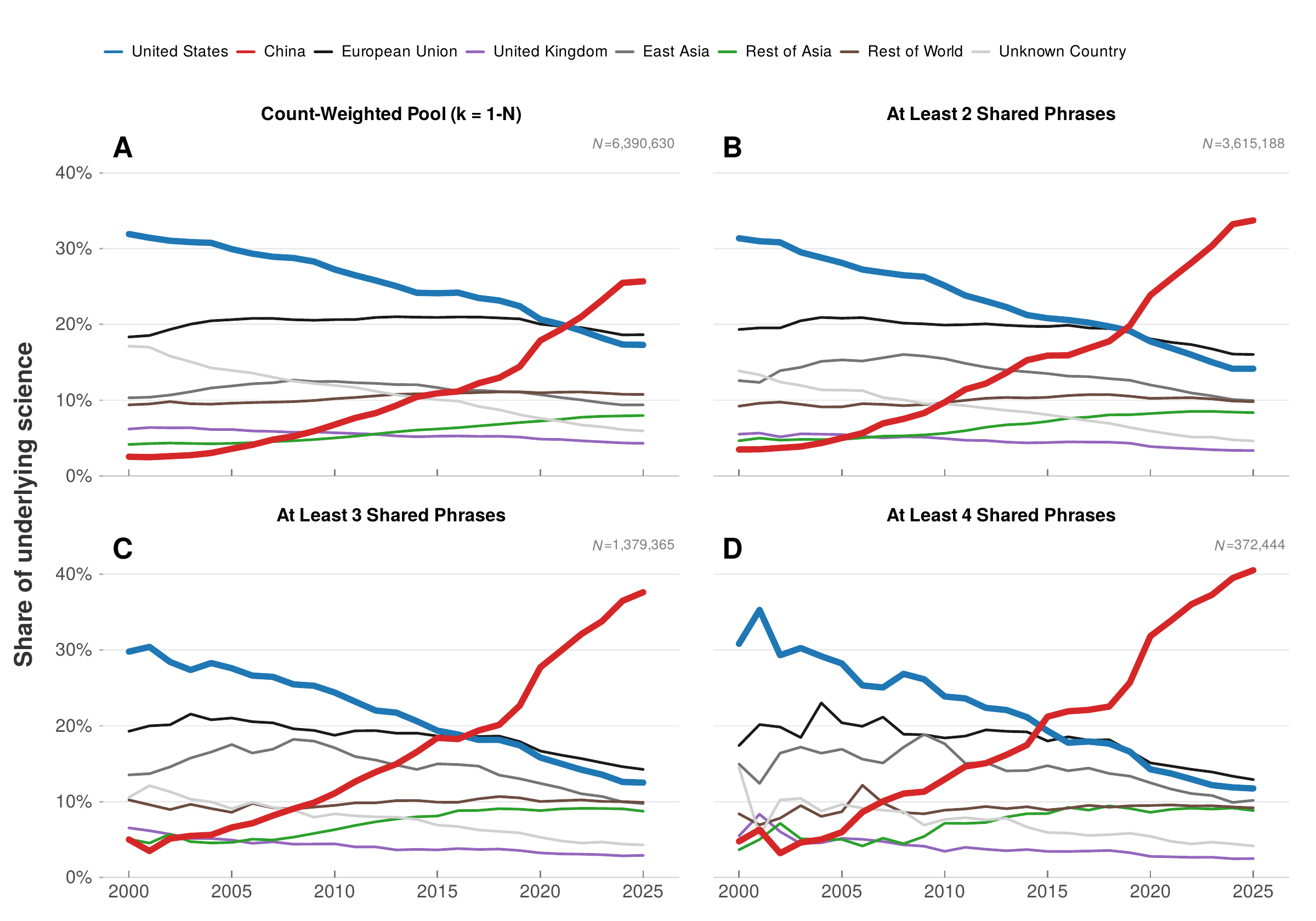}
	\caption{\textbf{Country shares of science underlying Chinese patents by phrase combination length.} (\textbf{A}) All tuples (of any length). (\textbf{B}) 2-tuples. (\textbf{C}) 3-tuples. (\textbf{D}) tuples of length 4 or greater.}
	\label{fig:combined_share_if1}
\end{figure}
\clearpage

\begin{figure}
	\centering
	\includegraphics[width=\textwidth]{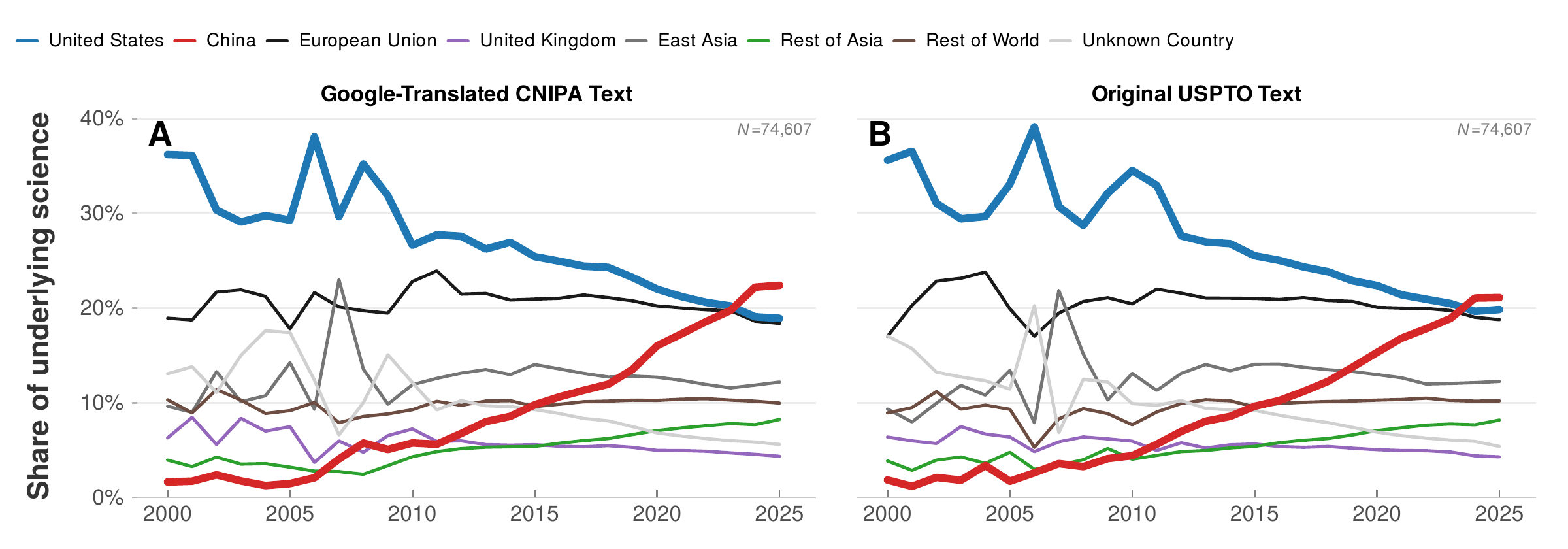}
	\caption{\textbf{Country shares of science underlying Chinese patents for patent sample filed at both the CNIPA and USPTO}. (\textbf{A}) Patent text supplied from CNIPA documents translated to English using Google Translate. (\textbf{B}) Patent text supplied by USPTO documents, which are submitted to the USPTO in English.}
	\label{fig:supp-translation}
\end{figure}

\clearpage
\begin{figure}
	\centering
	\includegraphics[width=\textwidth,height=0.82\textheight,keepaspectratio]{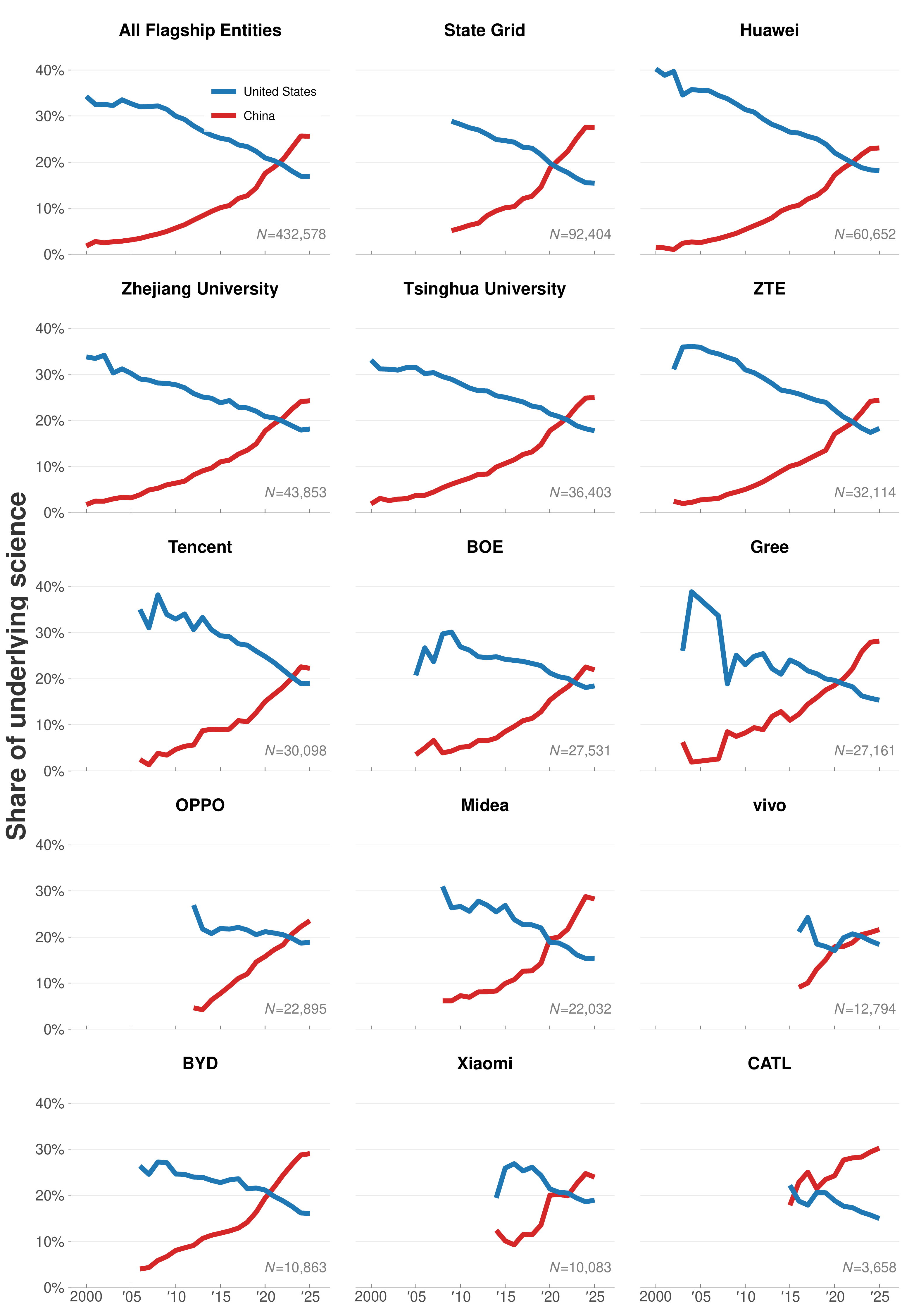}
	\caption{\textbf{Science underlying Chinese patents produced in the U.S. and China across flagship Chinese entities.} Each panel reports the share of linked scientific exposure from the United States and China for a flagship company or technical university. The first panel aggregates all flagship entities, and panel annotations report the number of CNIPA patents contributing to each estimate.}
	\label{fig:flagship-overview}
\end{figure}


\begin{figure}
	\centering
	\includegraphics[width=\textwidth]{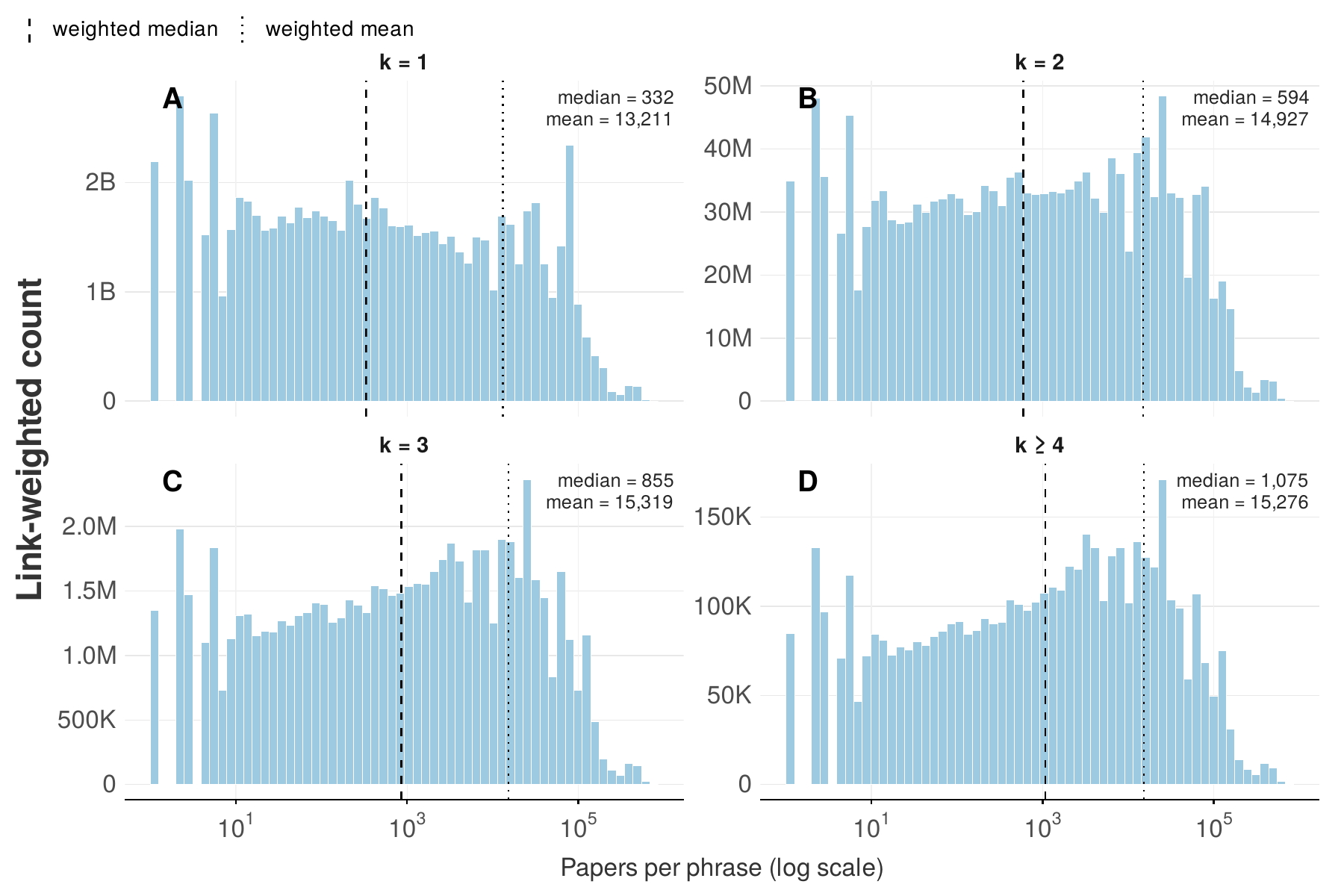}
	\caption{\textbf{Ubiquity of linked phrases.} Ubiquity of linked phrases, measured as the number of prior publications in which each phrase appears. (\textbf{A}) 1-tuples, (\textbf{B}) 2-tuples, (\textbf{C}) 3-tuples, (\textbf{D}) tuples of size 4 or larger.}
	\label{fig:supp-ubiquity}
\end{figure}

\clearpage


\clearpage
\input{cta_llm_prompt.tex}
\label{table:llm-prompt}
\clearpage

\clearpage
\begin{table}[t]
\centering
\small
\caption{\textbf{Character strings used to link CNIPA patents to Chinese organizations}}
\label{tab:cn-org-matching}
\begin{tabular}{@{}lp{6.0cm}p{2.3cm}r@{}}
\hline
Organization & Inclusion strings & Exclusion strings & $N$ \\
\hline
Huawei & \zh{华为}, \zh{海思半导体}, \zh{海思光电}, ``huawei'', ``hisilicon'' & --- & 80,758 \\
Tencent & \zh{腾讯}, ``tencent'' & --- & 37,010 \\
ZTE & \zh{中兴通讯}, \zh{中兴微电子}, \zh{中兴新软件}, \zh{中兴软件}, \zh{中兴力维}, \zh{中兴网信}, \zh{中兴派能}, ``zte corp'' & --- & 35,353 \\
BOE & \zh{京东方} & \zh{北京东方}, \zh{南京东方} & 32,108 \\
Xiaomi & \zh{小米}, ``xiaomi'' & --- & 14,398 \\
OPPO & \zh{欧珀}, ``oppo'' & --- & 27,882 \\
vivo & \zh{维沃} & --- & 15,641 \\
BYD & \zh{比亚迪} & --- & 12,167 \\
CATL & \zh{宁德时代}, ``contemporary amperex'' & --- & 5,070 \\
Gree & \zh{格力} & \zh{格力博} & 32,954 \\
Midea & \zh{美的} & --- & 24,924 \\
State Grid & \zh{国家电网}, \zh{国网}, ``state grid'' & \zh{中国网} & 102,207 \\
Tsinghua University & \zh{清华大学} & --- & 39,800 \\
Zhejiang University & \zh{浙江大学} & --- & 48,021 \\
\hline
\end{tabular}
\end{table}
\clearpage

\input{cta_class_table.tex}

\end{document}

%% file: cta_llm_prompt.tex
\clearpage
\refstepcounter{figure}\label{fig:cta-prompt}%
\begingroup
\scriptsize
\setlength{\parindent}{0pt}
\setlength{\parskip}{3pt}
\raggedright
\begin{framed}
{\scriptsize\textsc{System prompt}}\par\vspace{1pt}\hrule\vspace{3pt}
You are classifying International Patent Classification (IPC) main-group codes into one of 11 Critical Technology Areas (CTAs) defined by the U.S. National Science Foundation, or `Not Applicable' if no CTA fits.\par
The CTAs and their associated keywords are:
\begin{list}{}{\setlength{\leftmargin}{1.3em}\setlength{\itemindent}{0pt}\setlength{\itemsep}{2pt}\setlength{\topsep}{2pt}\setlength{\parsep}{0pt}}
\item[\textbf{1.}] \textbf{Advanced Communications} --- \textit{Keywords:} advanced communications technology, communications technology, impacts to communications from long term changes in atmospheric condition, wireless communication, optical communication, 5G, internet of things, satellite communication, network protocol, digital signal processing, telecommunication system, mobile network, broadband technologies, data transmission, communication security, machine to machine communication, radio frequency technologies, cognitive radio network, VOIP, multimedia communication, network architecture, cyber physical system, immersive technology, virtual reality, augmented reality, mixed reality, extended reality, 360 degree video, haptic feedback, immersive experience, spatial computing, virtual environment, wearable technology, head mounted display, 3D modeling, virtual world, virtual presence
\item[\textbf{2.}] \textbf{Artificial Intelligence} --- \textit{Keywords:} zero shot learning, artificial general intelligence, large language model, transfer learning, foundation model, deep reinforcement learning, human in the loop deep reinforcement learning, generative adversarial network, neural network, generative artificial intelligence, convolutional neural network, recurrent neural network, tensor processing unit, artificial neural network, ethics and responsible artificial intelligence, few shot learning, explainable artificial intelligence, one shot learning, deep learning, backpropagation, model evaluation, machine learning, adversarial machine learning, decision tree, random forest, support vector machine, regression, ensemble method, human machine teaming, robotics, unmanned system, autonomous system, sensor fusion, path planning, control system, human robot interaction, navigation system, supervised learning, unsupervised learning, semi supervised learning, multi agent system, natural language processing, data science, data analytics, data mining, classification, gradient descent, clustering, feature extraction, dimensionality reduction, hyperparameter tuning, computer vision
\item[\textbf{3.}] \textbf{Biotechnology} --- \textit{Keywords:} biotechnology, bioeconomy, medical technology, biomanufacturing, genomics, bioremediation, synthetic biology, genetic sequencing, biopharmaceutical, biomechanics, tissue engineering, biomaterial, regenerative medicine, biomedical imaging, medical device, drug delivery, genetic engineering, bioinformatics, molecular biology, cell therapy, nanotechnology, bioprocessing, bioreactor, prosthetic, orthopedic engineering, bioengineering, biocompatibility, biosensor, neural engineering, immunotherapy, stem cell, biomedical signal processing, bioprinting, biomem, biomedical microelectromechanical system, precision medicine, cardiovascular engineering, clinical engineering, biophotonic, bioelectrical engineering, medical robotics, biomedical data analysis, pharmacokinetics, biomedical optics, biomolecular engineering, biochemical engineering, biomedical diagnostics, medical imaging technique, biostatistics, microfluidics, biofabrication, personalized medicine, biomedical instrumentation, functional genomics, computational biology, tissue mechanics, rehabilitation engineering, biomedical ethics, biosignal processing
\item[\textbf{4.}] \textbf{Data Management and Security} --- \textit{Keywords:} data management, data governance, data quality, database system, data security, data warehousing, data integration, metadata management, information lifecycle management, data privacy, data architecture, master data management, data lake, data virtualization, NoSQL database, SQL, data cleaning, data migration, ETL, extract, transform, load, distributed database, real time data processing, data lifecycle management, cloud hosted database, database management system, data integrity, secure data storage, data backup and recovery, cybersecurity, encryption, network security, information security, cryptography, intrusion detection, firewall, malware analysis, risk management, access control, secure data transmission, phishing attack, data breach, blockchain, security protocol, threat intelligence, penetration testing, vulnerability assessment, cloud security, biometric security, identity management, secure authentication, compliance and regulation, advanced persistent threat, anomaly detection, security architecture, digital forensic, two factor authentication, zero trust security
\item[\textbf{5.}] \textbf{Disaster Resilience} --- \textit{Keywords:} anthropogenic disaster prevention, climate resilience and adaptation, anthropogenic disaster mitigation, climate resiliency, natural disaster prevention, climate adaptation, natural disaster mitigation, disaster prevention, disaster mitigation, climate change, environmental degradation, industrial accident, pollution, deforestation, oil spill, nuclear accident, hazardous waste, urbanization, land degradation, air pollution, water contamination, soil erosion, overfishing, biodiversity loss, chemical spill, mining disaster, greenhouse gas emission, toxic waste, ozone depletion, drought, desertification, environmental health, resource depletion, ecosystem destruction, global warming, acid rain, pesticide contamination, overpopulation, habitat loss, wildfire, carbon footprint, e-waste, landfill, ocean acidification, agricultural runoff, noise pollution, light pollution, plastic pollution, industrial emission, coral reef degradation, electromagnetic pollution, urban sprawl, invasive species, non-renewable energy, heavy metal contamination, environmental policy, ecological collapse, anthropogenic climate change
\item[\textbf{6.}] \textbf{Energy} --- \textit{Keywords:} energy efficiency technology, battery management control, industrial efficiency technology, battery system integration, batteries, nuclear technology, battery management system, renewable energy, solar power, wind energy, energy storage, smart grid, energy efficiency, bioenergy, photovoltaic, battery technology, nuclear energy, hydroelectric power, thermal energy, electric vehicle, energy policy, fossil fuel, greenhouse gas emission, sustainable energy, energy conversion, geothermal energy, energy management, power generation, energy market, carbon capture, energy harvesting, offshore wind, grid integration, energy demand, hydrogen fuel cell, power system, climate change, energy security, energy modeling, oil and gas, distributed energy resource, energy economics, biomass, energy innovation, solar cell, wave energy, energy transition, environmental impact, clean energy, demand response, energy conservation, smart meter, tidal energy, energy regulation, energy auditing, power electronics, energy sector
\item[\textbf{7.}] \textbf{High Performance Computing} --- \textit{Keywords:} high performance computing, parallel processing, high throughput computing, petascale grid computing, cluster computing, floating point operations per second, supercomputer, distributed computing, supercomputing, multi-core processing, load balancing, GPU computing, SIMD, MIMD, MPI, message passing interface, OpenMP, CUDA, memory management, data intensive computing, cloud computing, heterogeneous computing, scheduling algorithm, scientific computing, grid computing, data parallelism, task parallelism, high performance networking, real time computing, exascale computing, storage system, file system, I/O performance, workflow management, parallel file system, advanced vector extension, AVX, computational biology
\item[\textbf{8.}] \textbf{Materials Science} --- \textit{Keywords:} materials science, composite 2D material, next generation material, sustainable material, carbon footprint, biodegradable material, nanomaterial, biomaterial, energy harvesting, organic electronics, lightweight material, photovoltaic, green building material, bio-based polymer, polymer, composite, metallurgy, ceramic, superconductor, magnetic material, thin film, crystallography, carbon nanotube, graphene, alloy development, material characterization, corrosion, surface engineering, piezoelectric material, thermoelectric material, conductive material, mechanical properties, polymer chemistry, nanoengineering, material synthesis, quantum dot, ferromagnetic material, materials testing, smart material, molecular dynamics, photonic crystal, electrochemistry, nanofabrication, coating, structural material, polymer science, materials processing, tribology, electronic material, optical material, amorphous material, nanocomposite, nanowire, materials modeling, thermal properties, self assembly, membrane science
\item[\textbf{9.}] \textbf{Quantum Tech} --- \textit{Keywords:} quantum information science, quantum computer, quantum sensor, quantum technologies, quantum mechanics, quantum entanglement, quantum algorithm, quantum bit, qubit, superposition, quantum computing, quantum cryptography, quantum communication, quantum error correction, quantum key distribution, quantum teleportation, quantum gate, quantum circuit, quantum complexity theory, quantum simulation, quantum annealing, quantum decoherence, Bell state, quantum Fourier transform, no-cloning theorem, quantum machine learning, quantum sensing, quantum metrology, quantum network, quantum state preparation, quantum information theory, quantum optics, quantum coherence, quantum control, quantum logic gate, quantum measurement, quantum noise, quantum programming language, quantum software, quantum hardware, quantum process tomography, quantum phase estimation, quantum chemistry, quantum topological state, quantum artificial intelligence, quantum optimization, quantum probability, quantum tunneling, quantum field theory, Bloch sphere, quantum thermodynamics, quantum nonlocality, quantum entropy, quantum dynamics, Heisenberg uncertainty principle
\item[\textbf{10.}] \textbf{Robotics and Advanced Manufacturing} --- \textit{Keywords:} manufacturing processes, lean manufacturing, additive manufacturing, industrial engineering, supply chain management, quality control, production planning, mass production, computer aided manufacturing, CAM, computer numerical control, CNC, robotics, 3D printing, process optimization, materials science, industrial robotics, smart manufacturing, Six Sigma, flexible manufacturing system, machining, assembly line, industrial IoT, Kaizen, just in time production, MRP, materials requirement planning, CAD/CAM integration, manufacturing execution system, precision engineering, sustainable manufacturing, injection molding, process control, computer aided design, eco-friendly manufacturing, semiconductor, semiconductor material, silicon, gallium arsenide, transistor, integrated circuit, doping, semiconductor fabrication, microelectronics, photovoltaic, diode, Moore's law, electron mobility, hole mobility, CMOS technology, bipolar junction transistor, field effect transistor, semiconductor laser, optoelectronics, wafer fabrication, thin film transistor, semiconductor device, carrier concentration, P-N junction
\item[\textbf{11.}] \textbf{Semiconductors} --- \textit{Keywords:} semiconductor heterostructure, silicon carbide, organic semiconductor, photodetector, MEMS, microelectromechanical system, VLSI, very large scale integration, semiconductor packaging, electronic band structure, compound semiconductor, semiconductor metrology, charge carrier transport, sputtering, thermal oxidation, electron beam lithography, spintronics, superconducting semiconductor, semiconductor reliability, photolithography
\item[\textbf{12.}] \textbf{Not Applicable} --- Use this when the IPC code does not clearly fit any CTA above.
\end{list}
\textbf{Classification rules:}
\begin{list}{$\bullet$}{\setlength{\leftmargin}{1.3em}\setlength{\itemsep}{1pt}\setlength{\topsep}{2pt}\setlength{\parsep}{0pt}}
\item Choose the single best-fitting CTA based on the IPC code title.
\item Be conservative: if the code is generic or does not clearly match, use `Not Applicable'.
\item Watch for false positives in Quantum Tech, especially optical communications, nanomaterials, and quantum-dot materials that are better classified elsewhere.
\item Watch for false positives in Artificial Intelligence, especially biological neural systems that are not computational AI.
\item Watch for broad Materials Science terms that are actually more specific to Energy, Semiconductors, Biotechnology, or Advanced Manufacturing.
\end{list}
Return only a JSON object matching this schema: \texttt{ipc\_group}, \texttt{cta}, \texttt{confidence}, \texttt{reasoning}. Reasoning must be one short sentence. (\texttt{cta} is constrained to the 11 CTA names or \texttt{Not Applicable}; \texttt{confidence} is an integer 1--10.)\par
\vspace{3pt}{\scriptsize\textsc{User message} (one per IPC code)}\par\vspace{1pt}\hrule\vspace{3pt}
\texttt{Classify this IPC code:}\par
\texttt{Code: \{ipc\_group\}}\par
\texttt{Title: \{full\_title\}}
\end{framed}
\endgroup
\noindent\textbf{Figure \thefigure.} \textbf{Prompt used to classify IPC codes into Critical Technology Areas with a large language model.}\par

%% file: cta_class_table.tex
\begingroup
\refstepcounter{table}\label{tab:cta-ipc-crosswalk}%
\edef\ctacrosswalktablenumber{\thetable}%
\noindent\textbf{Table \ctacrosswalktablenumber.} \textbf{Crosswalk between Critical Technology Areas (CTAs) and IPC main-group codes}\par\vspace{4pt}
\scriptsize
\linespread{1}\selectfont
\setlength{\tabcolsep}{4pt}
\renewcommand{\arraystretch}{1.05}
\begin{longtable}{@{}p{0.205\textwidth}@{\hspace{6pt}}p{0.755\textwidth}@{}}
\hline
\textbf{Critical Technology Area} & \textbf{IPC main-group codes}\\
\hline
\endfirsthead
\multicolumn{2}{@{}l}{\emph{Table~\ctacrosswalktablenumber\ (continued)}}\\[2pt]
\hline
\textbf{Critical Technology Area} & \textbf{IPC main-group codes}\\
\hline
\endhead
\hline
\multicolumn{2}{r@{}}{\emph{continued on next page}}\\
\endfoot
\hline
\endlastfoot
\textbf{Advanced Communications (149)} & B61L 27/00, G01S 1/00, G01S 3/00, G01S 5/00, G01S 7/00, G01S 11/00, G01S 13/00, G01S 19/00, G02B 6/00, G02B 30/00, G02F 1/00, G02F 2/00, G03B 35/00, G06T 17/00, G06T 19/00, G08C 15/00, G08C 17/00, G08C 19/00, G16Y 10/00, G16Y 20/00, G16Y 30/00, G16Y 40/00, H01B 11/00, H01P 1/00, H01P 3/00, H01P 5/00, H01P 9/00, H01Q 1/00, H01Q 3/00, H01Q 5/00, H01Q 7/00, H01Q 9/00, H01Q 11/00, H01Q 13/00, H01Q 15/00, H01Q 17/00, H01Q 19/00, H01Q 21/00, H01Q 23/00, H01Q 25/00, H01S 1/00, H01S 4/00, H03C 1/00, H03C 3/00, H03C 5/00, H03C 7/00, H03C 99/00, H03D 1/00, H03D 3/00, H03D 5/00, H03D 7/00, H03D 9/00, H03D 11/00, H03D 13/00, H03D 99/00, H03K 7/00, H03M 3/00, H03M 13/00, H04B 1/00, H04B 3/00, H04B 5/00, H04B 7/00, H04B 10/00, H04B 11/00, H04B 13/00, H04B 14/00, H04B 15/00, H04B 17/00, H04H 20/00, H04H 40/00, H04H 60/00, H04J 1/00, H04J 3/00, H04J 4/00, H04J 7/00, H04J 9/00, H04J 11/00, H04J 13/00, H04J 14/00, H04J 99/00, H04K 3/00, H04L 1/00, H04L 5/00, H04L 7/00, H04L 12/00, H04L 13/00, H04L 17/00, H04L 23/00, H04L 25/00, H04L 27/00, H04L 41/00, H04L 43/00, H04L 45/00, H04L 47/00, H04L 49/00, H04L 51/00, H04L 61/00, H04L 65/00, H04L 67/00, H04L 69/00, H04L 101/00, H04M 1/00, H04M 3/00, H04M 7/00, H04M 9/00, H04M 11/00, H04M 13/00, H04M 15/00, H04M 99/00, H04N 1/00, H04N 5/00, H04N 7/00, H04N 11/00, H04N 13/00, H04N 19/00, H04N 21/00, H04Q 1/00, H04Q 3/00, H04Q 5/00, H04Q 9/00, H04Q 11/00, H04R 5/00, H04R 27/00, H04S 1/00, H04S 3/00, H04S 5/00, H04S 7/00, H04W 4/00, H04W 8/00, H04W 12/00, H04W 16/00, H04W 24/00, H04W 28/00, H04W 36/00, H04W 40/00, H04W 48/00, H04W 52/00, H04W 56/00, H04W 60/00, H04W 64/00, H04W 68/00, H04W 72/00, H04W 74/00, H04W 76/00, H04W 80/00, H04W 84/00, H04W 88/00, H04W 92/00, H04W 99/00\\[3pt]
\textbf{Artificial Intelligence (45)} & B60W 10/00, B60W 30/00, B60W 40/00, B60W 50/00, B60W 60/00, B62D 6/00, B62D 109/00, B64U 10/00, B64U 70/00, B64U 101/00, E21B 44/00, F41G 9/00, G01C 21/00, G01S 17/00, G05B 13/00, G05D 1/00, G05D 111/00, G06F 18/00, G06F 40/00, G06N 3/00, G06N 5/00, G06N 7/00, G06N 20/00, G06N 99/00, G06T 1/00, G06T 3/00, G06T 5/00, G06T 7/00, G06T 9/00, G06T 11/00, G06T 12/00, G06T 15/00, G06V 10/00, G06V 20/00, G06V 30/00, G06V 40/00, G08B 31/00, G10L 13/00, G10L 15/00, G10L 17/00, G10L 19/00, G10L 21/00, G10L 25/00, G16C 10/00, G16H 50/00\\[3pt]
\textbf{Biotechnology (200)} & A01H 1/00, A01H 3/00, A01H 4/00, A01K 67/00, A01N 1/00, A01N 63/00, A61B 1/00, A61B 3/00, A61B 5/00, A61B 6/00, A61B 7/00, A61B 8/00, A61B 10/00, A61B 17/00, A61B 18/00, A61B 34/00, A61B 50/00, A61B 90/00, A61C 3/00, A61C 8/00, A61C 13/00, A61C 19/00, A61D 7/00, A61D 17/00, A61D 19/00, A61F 2/00, A61F 3/00, A61F 4/00, A61F 5/00, A61F 7/00, A61F 9/00, A61F 11/00, A61H 1/00, A61H 3/00, A61H 31/00, A61J 1/00, A61J 3/00, A61J 7/00, A61J 15/00, A61K 9/00, A61K 31/00, A61K 33/00, A61K 35/00, A61K 38/00, A61K 39/00, A61K 40/00, A61K 41/00, A61K 45/00, A61K 47/00, A61K 48/00, A61K 49/00, A61K 50/00, A61K 51/00, A61K 101/00, A61K 103/00, A61K 129/00, A61L 2/00, A61L 12/00, A61L 15/00, A61L 17/00, A61L 24/00, A61L 26/00, A61L 27/00, A61L 28/00, A61L 29/00, A61L 31/00, A61L 33/00, A61L 101/00, A61L 103/00, A61M 1/00, A61M 3/00, A61M 5/00, A61M 11/00, A61M 13/00, A61M 15/00, A61M 16/00, A61M 19/00, A61M 21/00, A61M 25/00, A61M 27/00, A61M 29/00, A61M 31/00, A61M 35/00, A61M 36/00, A61M 37/00, A61M 39/00, A61M 60/00, A61M 99/00, A61N 1/00, A61N 2/00, A61N 5/00, A61N 7/00, A61P 1/00, A61P 3/00, A61P 5/00, A61P 7/00, A61P 9/00, A61P 11/00, A61P 13/00, A61P 15/00, A61P 17/00, A61P 19/00, A61P 21/00, A61P 23/00, A61P 25/00, A61P 27/00, A61P 29/00, A61P 31/00, A61P 33/00, A61P 35/00, A61P 37/00, A61P 39/00, A61P 41/00, A61P 43/00, B82Y 5/00, B82Y 15/00, C07C 405/00, C07D 477/00, C07D 499/00, C07D 501/00, C07D 503/00, C07D 505/00, C07G 11/00, C07H 19/00, C07H 21/00, C07K 1/00, C07K 2/00, C07K 4/00, C07K 5/00, C07K 7/00, C07K 9/00, C07K 11/00, C07K 14/00, C07K 16/00, C07K 17/00, C07K 19/00, C12M 1/00, C12M 3/00, C12N 1/00, C12N 3/00, C12N 5/00, C12N 7/00, C12N 9/00, C12N 11/00, C12N 13/00, C12N 15/00, C12P 1/00, C12P 3/00, C12P 5/00, C12P 7/00, C12P 9/00, C12P 11/00, C12P 13/00, C12P 15/00, C12P 17/00, C12P 19/00, C12P 21/00, C12P 23/00, C12P 25/00, C12P 27/00, C12P 29/00, C12P 31/00, C12P 33/00, C12P 35/00, C12P 37/00, C12P 39/00, C12P 41/00, C12Q 1/00, C12Q 3/00, C12R 1/00, C40B 10/00, C40B 20/00, C40B 30/00, C40B 40/00, C40B 50/00, C40B 60/00, C40B 70/00, C40B 80/00, D06M 16/00, G01N 33/00, G01N 35/00, G16B 5/00, G16B 10/00, G16B 15/00, G16B 20/00, G16B 25/00, G16B 30/00, G16B 35/00, G16B 40/00, G16B 45/00, G16B 50/00, G16B 99/00, G16H 10/00, G16H 15/00, G16H 20/00, G16H 30/00, G16H 40/00, G16H 80/00, G21K 4/00, H04R 25/00\\[3pt]
\textbf{Data Management and Security (22)} & B42D 25/00, B60R 25/00, E05B 47/00, E05B 49/00, E05B 51/00, G06F 11/00, G06F 16/00, G06F 21/00, G06Q 20/00, G07C 9/00, G08B 29/00, G09C 1/00, G09C 3/00, G09C 5/00, G11B 20/00, G11C 7/00, G11C 15/00, G11C 29/00, G16C 20/00, G16H 70/00, H04K 1/00, H04L 9/00\\[3pt]
\textbf{Disaster Resilience (15)} & A01G 15/00, A62D 3/00, A62D 101/00, B01D 53/00, B09B 1/00, B09B 3/00, B09C 1/00, C02F 101/00, C09K 105/00, E01F 7/00, E04H 9/00, E21F 5/00, F01N 3/00, F23J 15/00, G21F 9/00\\[3pt]
\textbf{Energy (170)} & B60K 1/00, B60K 6/00, B60K 16/00, B60L 1/00, B60L 3/00, B60L 5/00, B60L 7/00, B60L 8/00, B60L 9/00, B60L 13/00, B60L 15/00, B60L 50/00, B60L 53/00, B60L 55/00, B60L 58/00, B60M 1/00, B60M 3/00, B60W 20/00, B61D 43/00, B62J 43/00, C01B 3/00, C09K 5/00, C10B 19/00, C10G 1/00, C10G 2/00, C10G 3/00, C10G 5/00, C10J 1/00, C10J 3/00, C10K 1/00, C10K 3/00, C10L 1/00, C10L 3/00, C10L 5/00, C10L 7/00, C10L 8/00, C10L 9/00, C10L 10/00, C25B 5/00, C25B 9/00, E02B 9/00, F01K 1/00, F01K 3/00, F01K 5/00, F01K 7/00, F01K 17/00, F01K 19/00, F01K 23/00, F01K 25/00, F01K 27/00, F01N 5/00, F02B 41/00, F02G 1/00, F02G 5/00, F03B 1/00, F03B 3/00, F03B 7/00, F03B 11/00, F03B 13/00, F03B 15/00, F03D 1/00, F03D 3/00, F03D 5/00, F03D 7/00, F03D 9/00, F03D 13/00, F03D 15/00, F03D 17/00, F03D 80/00, F03G 4/00, F03G 6/00, F16D 61/00, F23L 15/00, F24D 11/00, F24D 13/00, F24D 18/00, F24D 101/00, F24D 103/00, F24D 105/00, F24F 12/00, F24H 4/00, F24S 10/00, F24S 20/00, F24S 21/00, F24S 23/00, F24S 25/00, F24S 30/00, F24S 40/00, F24S 50/00, F24S 60/00, F24S 70/00, F24S 80/00, F24S 90/00, F24T 10/00, F24T 50/00, F24V 50/00, F25B 27/00, F25B 29/00, F25B 30/00, F25D 16/00, F27D 17/00, F28D 20/00, G21B 1/00, G21B 3/00, G21C 1/00, G21C 3/00, G21C 5/00, G21C 7/00, G21C 9/00, G21C 11/00, G21C 13/00, G21C 15/00, G21C 17/00, G21C 19/00, G21C 21/00, G21C 23/00, G21D 1/00, G21D 3/00, G21D 5/00, G21D 7/00, G21D 9/00, G21F 1/00, G21F 7/00, G21G 1/00, G21G 4/00, G21H 1/00, G21H 3/00, G21J 1/00, H01G 11/00, H01M 4/00, H01M 6/00, H01M 8/00, H01M 10/00, H01M 12/00, H01M 14/00, H01M 16/00, H01M 50/00, H02H 7/00, H02J 1/00, H02J 3/00, H02J 4/00, H02J 7/00, H02J 9/00, H02J 11/00, H02J 13/00, H02J 15/00, H02J 50/00, H02J 101/00, H02J 103/00, H02J 105/00, H02J 107/00, H02K 35/00, H02K 44/00, H02M 1/00, H02M 3/00, H02M 5/00, H02M 7/00, H02M 9/00, H02M 11/00, H02N 3/00, H02N 10/00, H02P 9/00, H02P 27/00, H02P 103/00, H02S 10/00, H02S 20/00, H02S 30/00, H02S 40/00, H02S 50/00, H02S 99/00\\[3pt]
\textbf{High Performance Computing (6)} & G06E 1/00, G06E 3/00, G06F 12/00, G06F 13/00, G06J 1/00, G06J 3/00\\[3pt]
\textbf{Quantum Tech (4)} & B82Y 10/00, G06N 10/00, H02K 55/00, H03F 19/00\\[3pt]
\textbf{Materials Science (377)} & B01D 71/00, B01J 35/00, B32B 1/00, B32B 3/00, B32B 5/00, B32B 7/00, B32B 9/00, B32B 13/00, B32B 15/00, B32B 17/00, B32B 18/00, B32B 19/00, B32B 23/00, B32B 25/00, B32B 27/00, B32B 33/00, B32B 37/00, B60C 1/00, B62D 29/00, B82B 1/00, B82B 3/00, B82Y 20/00, B82Y 25/00, B82Y 30/00, B82Y 35/00, B82Y 40/00, C01B 32/00, C01G 23/00, C01G 35/00, C03C 1/00, C03C 3/00, C03C 4/00, C03C 6/00, C03C 8/00, C03C 10/00, C03C 11/00, C03C 12/00, C03C 13/00, C03C 14/00, C03C 15/00, C03C 17/00, C03C 19/00, C03C 21/00, C03C 23/00, C03C 25/00, C03C 29/00, C04B 5/00, C04B 14/00, C04B 16/00, C04B 18/00, C04B 20/00, C04B 24/00, C04B 26/00, C04B 30/00, C04B 32/00, C04B 33/00, C04B 35/00, C04B 38/00, C04B 41/00, C04B 103/00, C08B 1/00, C08B 3/00, C08B 5/00, C08B 7/00, C08B 9/00, C08B 11/00, C08B 13/00, C08B 15/00, C08B 16/00, C08B 31/00, C08B 33/00, C08B 35/00, C08B 37/00, C08C 1/00, C08C 2/00, C08C 3/00, C08C 4/00, C08C 19/00, C08F 2/00, C08F 4/00, C08F 6/00, C08F 8/00, C08F 10/00, C08F 12/00, C08F 14/00, C08F 16/00, C08F 18/00, C08F 20/00, C08F 22/00, C08F 24/00, C08F 26/00, C08F 28/00, C08F 30/00, C08F 32/00, C08F 34/00, C08F 36/00, C08F 38/00, C08F 110/00, C08F 112/00, C08F 114/00, C08F 116/00, C08F 118/00, C08F 120/00, C08F 122/00, C08F 124/00, C08F 126/00, C08F 128/00, C08F 130/00, C08F 132/00, C08F 134/00, C08F 136/00, C08F 138/00, C08F 210/00, C08F 212/00, C08F 214/00, C08F 216/00, C08F 218/00, C08F 220/00, C08F 222/00, C08F 224/00, C08F 226/00, C08F 228/00, C08F 230/00, C08F 232/00, C08F 234/00, C08F 236/00, C08F 238/00, C08F 240/00, C08F 242/00, C08F 244/00, C08F 246/00, C08F 251/00, C08F 253/00, C08F 255/00, C08F 257/00, C08F 259/00, C08F 261/00, C08F 263/00, C08F 265/00, C08F 267/00, C08F 269/00, C08F 271/00, C08F 273/00, C08F 275/00, C08F 277/00, C08F 279/00, C08F 281/00, C08F 283/00, C08F 285/00, C08F 287/00, C08F 289/00, C08F 290/00, C08F 291/00, C08F 292/00, C08F 293/00, C08F 295/00, C08F 297/00, C08F 299/00, C08F 301/00, C08G 2/00, C08G 4/00, C08G 8/00, C08G 10/00, C08G 12/00, C08G 14/00, C08G 16/00, C08G 18/00, C08G 59/00, C08G 61/00, C08G 63/00, C08G 64/00, C08G 65/00, C08G 67/00, C08G 69/00, C08G 71/00, C08G 73/00, C08G 75/00, C08G 77/00, C08G 79/00, C08G 81/00, C08G 83/00, C08G 85/00, C08G 101/00, C08H 1/00, C08H 7/00, C08H 8/00, C08J 3/00, C08J 5/00, C08J 7/00, C08J 9/00, C08K 3/00, C08K 5/00, C08K 7/00, C08K 9/00, C08K 13/00, C08L 1/00, C08L 3/00, C08L 5/00, C08L 7/00, C08L 9/00, C08L 11/00, C08L 13/00, C08L 15/00, C08L 17/00, C08L 19/00, C08L 21/00, C08L 23/00, C08L 25/00, C08L 27/00, C08L 29/00, C08L 31/00, C08L 33/00, C08L 35/00, C08L 37/00, C08L 39/00, C08L 41/00, C08L 43/00, C08L 45/00, C08L 47/00, C08L 49/00, C08L 51/00, C08L 53/00, C08L 55/00, C08L 57/00, C08L 59/00, C08L 61/00, C08L 63/00, C08L 65/00, C08L 67/00, C08L 69/00, C08L 71/00, C08L 73/00, C08L 75/00, C08L 77/00, C08L 79/00, C08L 81/00, C08L 83/00, C08L 85/00, C08L 87/00, C08L 89/00, C08L 93/00, C08L 97/00, C08L 99/00, C08L 101/00, C09J 101/00, C09J 181/00, C09J 183/00, C09K 9/00, C09K 11/00, C09K 19/00, C09K 111/00, C09K 113/00, C21D 1/00, C21D 3/00, C21D 5/00, C21D 6/00, C21D 7/00, C21D 8/00, C21D 9/00, C21D 10/00, C22B 1/00, C22B 5/00, C22B 7/00, C22B 9/00, C22B 11/00, C22B 21/00, C22B 34/00, C22B 35/00, C22B 59/00, C22C 1/00, C22C 3/00, C22C 5/00, C22C 7/00, C22C 9/00, C22C 11/00, C22C 12/00, C22C 13/00, C22C 14/00, C22C 16/00, C22C 18/00, C22C 19/00, C22C 20/00, C22C 21/00, C22C 22/00, C22C 23/00, C22C 24/00, C22C 25/00, C22C 26/00, C22C 27/00, C22C 28/00, C22C 29/00, C22C 30/00, C22C 32/00, C22C 33/00, C22C 35/00, C22C 37/00, C22C 38/00, C22C 43/00, C22C 45/00, C22C 47/00, C22C 49/00, C22C 101/00, C22C 111/00, C22C 121/00, C22F 1/00, C22F 3/00, C23C 2/00, C23C 4/00, C23C 6/00, C23C 8/00, C23C 10/00, C23C 12/00, C23C 14/00, C23C 18/00, C23C 20/00, C23C 22/00, C23C 24/00, C23C 26/00, C23C 28/00, C23C 30/00, C23D 3/00, C23D 5/00, C23D 7/00, C23F 1/00, C23F 3/00, C23F 4/00, C23F 11/00, C23F 13/00, C23F 15/00, C23F 17/00, C25D 3/00, C25D 9/00, C25D 11/00, C25D 15/00, C30B 1/00, C30B 3/00, C30B 5/00, C30B 7/00, C30B 9/00, C30B 11/00, C30B 13/00, C30B 17/00, C30B 21/00, C30B 23/00, C30B 27/00, C30B 28/00, C30B 29/00, C30B 30/00, C30B 33/00, C30B 35/00, D01F 1/00, D01F 2/00, D01F 6/00, D01F 8/00, D01F 9/00, D01F 11/00, D06M 14/00, F28F 21/00, G01N 3/00, G01N 13/00, G01N 17/00, G01N 19/00, G01N 23/00, G01Q 60/00, G01Q 70/00, G02B 1/00, G16C 60/00, H01B 1/00, H01B 3/00, H01B 12/00, H01F 1/00, H01F 3/00, H01F 6/00, H01F 10/00, H01F 36/00, H02N 2/00, H10N 30/00\\[3pt]
\textbf{Robotics and Advanced Manufacturing (338)} & A01B 69/00, A41H 42/00, B21B 1/00, B21B 5/00, B21B 9/00, B21B 15/00, B21B 17/00, B21B 19/00, B21B 25/00, B21B 37/00, B21B 38/00, B21B 39/00, B21B 47/00, B21C 23/00, B21C 31/00, B21C 33/00, B21C 37/00, B21C 51/00, B21D 1/00, B21D 3/00, B21D 5/00, B21D 7/00, B21D 9/00, B21D 11/00, B21D 13/00, B21D 15/00, B21D 17/00, B21D 19/00, B21D 21/00, B21D 22/00, B21D 24/00, B21D 25/00, B21D 26/00, B21D 28/00, B21D 31/00, B21D 33/00, B21D 35/00, B21D 37/00, B21D 39/00, B21D 41/00, B21D 43/00, B21D 45/00, B21D 47/00, B21D 49/00, B21D 51/00, B21D 53/00, B21F 5/00, B21F 31/00, B21F 37/00, B21H 9/00, B21J 1/00, B21J 3/00, B21J 5/00, B21J 9/00, B21J 11/00, B21J 13/00, B21J 15/00, B21J 17/00, B21K 1/00, B21K 3/00, B21K 5/00, B21K 25/00, B21K 27/00, B21K 31/00, B22C 1/00, B22C 3/00, B22C 5/00, B22C 7/00, B22C 9/00, B22C 11/00, B22C 13/00, B22C 15/00, B22C 17/00, B22C 19/00, B22C 21/00, B22C 23/00, B22C 25/00, B22D 1/00, B22D 2/00, B22D 7/00, B22D 9/00, B22D 11/00, B22D 13/00, B22D 15/00, B22D 17/00, B22D 18/00, B22D 19/00, B22D 21/00, B22D 23/00, B22D 25/00, B22D 27/00, B22D 29/00, B22D 30/00, B22D 31/00, B22D 33/00, B22D 35/00, B22D 37/00, B22D 39/00, B22D 43/00, B22D 45/00, B22D 46/00, B22D 47/00, B22F 1/00, B22F 3/00, B22F 5/00, B22F 7/00, B22F 8/00, B22F 9/00, B22F 10/00, B22F 12/00, B23B 1/00, B23B 3/00, B23B 5/00, B23B 7/00, B23B 9/00, B23B 11/00, B23B 13/00, B23B 15/00, B23B 17/00, B23B 19/00, B23B 21/00, B23B 25/00, B23B 27/00, B23B 29/00, B23B 31/00, B23B 33/00, B23B 35/00, B23B 37/00, B23B 39/00, B23B 41/00, B23B 43/00, B23B 45/00, B23B 47/00, B23B 49/00, B23B 51/00, B23C 1/00, B23C 3/00, B23C 5/00, B23C 7/00, B23C 9/00, B23D 1/00, B23D 3/00, B23D 5/00, B23D 7/00, B23D 11/00, B23D 13/00, B23D 15/00, B23D 17/00, B23D 19/00, B23D 21/00, B23D 23/00, B23D 25/00, B23D 27/00, B23D 31/00, B23D 33/00, B23D 35/00, B23D 36/00, B23D 37/00, B23D 39/00, B23D 41/00, B23D 43/00, B23D 45/00, B23D 47/00, B23D 49/00, B23D 51/00, B23D 53/00, B23D 55/00, B23D 57/00, B23D 59/00, B23D 61/00, B23D 63/00, B23D 65/00, B23D 67/00, B23D 69/00, B23D 75/00, B23D 77/00, B23D 79/00, B23D 81/00, B23F 1/00, B23F 3/00, B23F 5/00, B23F 7/00, B23F 9/00, B23F 11/00, B23F 13/00, B23F 15/00, B23F 17/00, B23F 19/00, B23F 21/00, B23F 23/00, B23G 1/00, B23G 3/00, B23G 5/00, B23G 7/00, B23G 9/00, B23G 11/00, B23H 1/00, B23H 3/00, B23H 5/00, B23H 7/00, B23H 9/00, B23H 11/00, B23K 1/00, B23K 3/00, B23K 5/00, B23K 7/00, B23K 9/00, B23K 10/00, B23K 11/00, B23K 13/00, B23K 15/00, B23K 20/00, B23K 25/00, B23K 26/00, B23K 28/00, B23K 31/00, B23K 33/00, B23K 35/00, B23K 37/00, B23K 101/00, B23K 103/00, B23P 6/00, B23P 9/00, B23P 11/00, B23P 13/00, B23P 15/00, B23P 17/00, B23P 19/00, B23P 21/00, B23P 23/00, B23P 25/00, B23Q 1/00, B23Q 3/00, B23Q 5/00, B23Q 7/00, B23Q 9/00, B23Q 11/00, B23Q 15/00, B23Q 16/00, B23Q 17/00, B23Q 23/00, B23Q 27/00, B23Q 33/00, B23Q 35/00, B23Q 37/00, B23Q 39/00, B23Q 41/00, B24B 1/00, B24B 5/00, B24B 7/00, B24B 9/00, B24B 11/00, B24B 17/00, B24B 29/00, B24B 31/00, B24B 33/00, B24B 35/00, B24B 37/00, B24B 39/00, B24B 41/00, B24B 49/00, B24B 51/00, B24B 53/00, B24B 57/00, B25J 1/00, B25J 3/00, B25J 5/00, B25J 7/00, B25J 9/00, B25J 11/00, B25J 13/00, B25J 15/00, B25J 17/00, B25J 18/00, B25J 19/00, B25J 21/00, B29B 11/00, B29C 31/00, B29C 33/00, B29C 39/00, B29C 41/00, B29C 43/00, B29C 45/00, B29C 48/00, B29C 51/00, B29C 53/00, B29C 55/00, B29C 59/00, B29C 64/00, B29C 65/00, B29C 67/00, B29C 69/00, B29C 70/00, B29K 227/00, B29K 263/00, B29K 269/00, B29K 301/00, B29K 303/00, B29K 307/00, B30B 11/00, B32B 39/00, B32B 41/00, B32B 43/00, B33Y 10/00, B33Y 30/00, B33Y 40/00, B33Y 50/00, B33Y 70/00, B33Y 80/00, B33Y 99/00, B62D 57/00, B62D 65/00, B64C 19/00, B64F 5/00, B64U 20/00, B64U 30/00, B64U 40/00, B65G 61/00, C08B 17/00, C21D 11/00, C25D 1/00, D05B 19/00, D05C 5/00, G05B 19/00, G05D 105/00, G06F 30/00, G06F 111/00, H01C 17/00, H01F 41/00, H01G 13/00, H01H 11/00, H01H 65/00, H01P 11/00, H01R 43/00, H02K 15/00, H03H 3/00, H05K 3/00, H05K 13/00\\[3pt]
\textbf{Semiconductors (143)} & B28D 5/00, B81B 1/00, B81B 3/00, B81B 5/00, B81B 7/00, B81C 1/00, B81C 3/00, B81C 99/00, C01B 33/00, C01G 15/00, C22B 58/00, C23C 16/00, C25F 3/00, C30B 15/00, C30B 19/00, C30B 25/00, C30B 31/00, F21K 9/00, F21Y 115/00, G03F 1/00, G03F 5/00, G03F 7/00, G03F 9/00, G11C 5/00, G11C 11/00, G11C 16/00, H01C 7/00, H01G 4/00, H01G 7/00, H01G 9/00, H01J 37/00, H01J 40/00, H01R 12/00, H01S 3/00, H01S 5/00, H02K 29/00, H03H 9/00, H03K 19/00, H04N 25/00, H05K 1/00, H10B 10/00, H10B 12/00, H10B 20/00, H10B 41/00, H10B 43/00, H10B 51/00, H10B 53/00, H10B 61/00, H10B 63/00, H10B 69/00, H10B 80/00, H10B 99/00, H10D 1/00, H10D 8/00, H10D 10/00, H10D 12/00, H10D 18/00, H10D 30/00, H10D 44/00, H10D 48/00, H10D 62/00, H10D 64/00, H10D 80/00, H10D 84/00, H10D 86/00, H10D 87/00, H10D 88/00, H10D 89/00, H10D 99/00, H10F 10/00, H10F 19/00, H10F 30/00, H10F 39/00, H10F 55/00, H10F 71/00, H10F 77/00, H10F 99/00, H10H 20/00, H10H 29/00, H10H 99/00, H10K 10/00, H10K 19/00, H10K 30/00, H10K 39/00, H10K 50/00, H10K 59/00, H10K 65/00, H10K 71/00, H10K 77/00, H10K 85/00, H10K 99/00, H10K 101/00, H10K 102/00, H10N 10/00, H10N 15/00, H10N 19/00, H10N 35/00, H10N 39/00, H10N 50/00, H10N 52/00, H10N 59/00, H10N 60/00, H10N 69/00, H10N 70/00, H10N 79/00, H10N 80/00, H10N 89/00, H10N 97/00, H10N 99/00, H10P 10/00, H10P 14/00, H10P 30/00, H10P 32/00, H10P 34/00, H10P 36/00, H10P 50/00, H10P 52/00, H10P 54/00, H10P 56/00, H10P 58/00, H10P 70/00, H10P 72/00, H10P 74/00, H10P 76/00, H10P 90/00, H10P 95/00, H10W 10/00, H10W 15/00, H10W 20/00, H10W 29/00, H10W 40/00, H10W 42/00, H10W 44/00, H10W 46/00, H10W 70/00, H10W 72/00, H10W 74/00, H10W 76/00, H10W 78/00, H10W 80/00, H10W 90/00, H10W 95/00, H10W 99/00\\[3pt]
\end{longtable}
\endgroup

%% file: bib_april_5_2026.bib
@article{fleming_technology_2001,
	title = {Technology as a complex adaptive system: evidence from patent data},
	author = {Fleming, Lee and Sorenson, Olav},
	year = {2001},
	journal = {Research Policy},
    Volume = {30},
	pages = {1019--1039},
}

@article{alshebli_china_2023,
	title = {China and the {U}.{S}. {Produce} more {Impactful} {AI} {Research} when {Collaborating} {Together}},
	url = {https://arxiv.org/abs/2304.11123},
	journal = {arXiv},
	author = {AlShebli, Bedoor and Ali Memon, Shahan and Evans, James and Rahwan, Talal},
	year = {2023},
}

@article{marx_reliance_2020,
	title = {Reliance on {Science}: {Worldwide} {Front}-{Page} {Patent} {Citations} to {Scientific} {Articles}},
	volume = {41},
	number = {9},
	journal = {2020},
	author = {Marx, Matt and Fuegi, Aaron},
	year = {2020},
	pages = {1572--1594},
}

@article{wu_chinas_2024,
	title = {Shifting power asymmetries in scientific teams reveal China’s rising leadership in global science},
	journal = {Proceedings of the National Academy of Science},
	author = {Wu, Renli and Esposito, Christopher and Evans, James},
	year = {2025},
    volume = {122},
	number = {44}
}

@incollection{chatterji_how_2025,
	series = {{NBER} {Book} {Series}},
	title = {How {Geopolitics} {Is} {Changing} the {Economics} of {Innovation}},
	volume = {5},
	booktitle = {Entrepreneurship and {Innovation} {Policy} and the {Economy}},
	publisher = {University of Chicago Press},
	author = {Chatterji, Aaron and Murray, Fiona},
	year = {2025},
}

@article{esposito_global_2026,
	title = {Global {Science} {Sustains} {U}.{S}. {Innovation}},
	journal = {Working Paper},
	author = {Esposito, Christopher},
	year = {2026},
}

@article{arts_beyond_2025,
	title = {Beyond {Citations}: {Measuring} {Novel} {Scientific} {Ideas} and their {Impact} in {Publication} {Text}},
	number = {Forthcoming},
	journal = {Review of Economics and Statistics},
	publisher = {MIT Press Journals},
	author = {Arts, Sam and Melluso, Nicola and Veugelers, Reinhilde},
	year = {2025},
	pages = {1--33},
}

@article{wang_chinese_2025,
	title = {Chinese firms are increasingly leveraging science for innovation, and geopolitical tensions have accelerated the process},
	journal = {Working Paper},
	author = {Wang, Junhan and Li, Xibao and Wang, Yanbo},
	year = {2025},
}

@misc{mervis_lawmakers_2025,
  author       = {Mervis, Jeffrey},
  title        = {Lawmakers Propose Banning All {U.S.}--{C}hinese Research Collaborations},
  howpublished = {Science (News)},
  publisher    = {American Association for the Advancement of Science},
  year         = {2026},
  month        = may,
  day          = {27},
  note         = {Accessed: 2026-05-28},
  url          = {https://www.science.org/content/article/lawmakers-propose-banning-all-u-s-chinese-research-collaborations}
}

@misc{sira_act_2026,
  author       = {{U.S. Congress}},
  title        = {Securing Innovation and Research from Adversaries Act},
  howpublished = {S.~4525, 119th Congress (2025--2026)},
  year         = {2026},
  note         = {Introduced by Sen.~J.~Banks and Rep.~J.~Moolenaar, May 2026},
  url          = {https://www.congress.gov/bill/119th-congress/senate-bill/4525}
}

@misc{chips_act_2022,
  author       = {{U.S. Congress}},
  title        = {{CHIPS} and Science Act of 2022},
  howpublished = {Public Law No.~117--167, 136 Stat.~1366},
  year         = {2022},
  month        = aug,
  day          = {9},
  note         = {H.R.~4346, 117th Congress},
  url          = {https://www.congress.gov/bill/117th-congress/house-bill/4346}
}

@misc{wagner_china_2026,
  author       = {Wagner, Caroline S.},
  title        = {China Surpasses the {US} in Research Spending: The Consequences
                  Extend Far Beyond Scientific Ranking and Clout},
  howpublished = {The Conversation},
  year         = {2026},
  month        = may,
  day          = {14},
  note         = {Reports OECD data showing China's R\&D spending reached parity
                  with and, by purchasing-power measures, surpassed that of the
                  United States, with both crossing \$1 trillion},
  url          = {https://theconversation.com/china-surpasses-us-in-research-spending-the-consequences-extend-far-beyond-scientific-ranking-and-clout-280543}
}

@article{kwon_dual_use_2026,
  author  = {Kwon, Seokbeom},
  title   = {Dual-Use Research under Scrutiny},
  journal = {Science},
  year    = {2026},
  month   = jun,
  day     = {4},
  volume  = {392},
  number  = {6802},
  pages   = {1032--1035},
  doi     = {10.1126/science.aee2479},
  url     = {https://www.science.org/doi/10.1126/science.aee2479}
}

@article{arts_melluso_2026,
  author  = {Arts, Sam and Melluso, Nicola},
  title   = {Hidden Reliance on Scientific Ideas and Firm Innovation},
  journal = {Working Paper},
  year    = {2026}}

@patent{hu_layered_2017,
  title       = {Layered copper-containing oxide material and preparation method and use thereof},
  author      = {Hu, Yongsheng and Li, Yunming and Xu, Shuyin and Wang, Yuesheng and Chen, Liquan and Huang, Xuejie},
  number      = {CN104795551B},
  nationality = {China},
  type        = {Invention patent},
  assignee    = {Institute of Physics, Chinese Academy of Sciences},
  year        = {2017},
  month       = jul,
  note        = {Priority date 2014-07-17; granted 2017-07-14},
  url         = {https://patents.google.com/patent/CN104795551B/en}
}
